\documentclass[12pt,letterpaper]{article}
\usepackage{amsfonts}
\usepackage{amssymb}
\usepackage{amsmath}
\usepackage{amsthm}
\usepackage[round,authoryear]{natbib}
\usepackage[margin=1in]{geometry}
\usepackage{booktabs}
\usepackage{dcolumn}
\newcolumntype{d}[1]{D{.}{.}{#1}}
\usepackage{graphicx,epstopdf}
\usepackage[T1]{fontenc}
\usepackage{lmodern}
\usepackage{enumitem}
\setlist[enumerate,1]{itemsep=1pt, topsep=1pt, partopsep=0pt, parsep=1pt}
\setlist[enumerate,2]{nosep}
\setlist[itemize,1]{itemsep=1pt, topsep=1pt, partopsep=0pt, parsep=1pt}
\setlist[itemize,2]{nosep}
\usepackage{subfig}
\usepackage{bm}
\usepackage[colorlinks,linkcolor=red,citecolor=blue,pdftex,breaklinks]{hyperref}
\hypersetup{bookmarksopen=true}
\usepackage[nameinlink]{cleveref}
\usepackage{setspace}

\theoremstyle{plain}

\theoremstyle{definition}

\Crefname{assumptionx}{Assumption}{Assumptions} 
\Crefname{examplex}{Example}{Examples} 
\Crefname{remarkx}{Remark}{Remarks} 

\makeatletter
\renewcommand*{\eqref}[1]{\hyperref[{#1}]{\textup{\tagform@{\ref*{#1}}}}}
\makeatother

\makeatletter
\DeclareRobustCommand\citepos
  {\begingroup\def\NAT@nmfmt##1{{\NAT@up##1's}}%
   \NAT@swafalse\let\NAT@ctype\z@\NAT@partrue
   \@ifstar{\NAT@fulltrue\NAT@citetp}{\NAT@fullfalse\NAT@citetp}}
\makeatother

\expandafter
\def \expandafter \normalsize \expandafter{\normalsize \setlength \abovedisplayskip{10pt plus 2pt minus 7pt}}
\expandafter
\def \expandafter \normalsize \expandafter{\normalsize \setlength \abovedisplayshortskip{0pt plus 2pt}}
\expandafter
\def \expandafter \normalsize \expandafter{\normalsize \setlength \belowdisplayskip{10pt plus 2pt minus 7pt}}
\expandafter
\def \expandafter \normalsize \expandafter{\normalsize \setlength \belowdisplayshortskip{5pt plus 2pt minus 3pt}}

\def\bia{{\bm{a}}}
\def\bbeta{{\bm\beta}}
\def\biX{{\bm{X}}}
\def\biZ{{\bm{Z}}}

\def\biM{{\bm{M}}}
\def\biP{{\bm{P}}}

\def\biH{{\bm{H}}}
\def\biR{{\bm{R}}}

\def\bir{{\bm{r}}}
\def\bis{{\bm{s}}}
\def\biy{{\bm{y}}}
\def\bix{{\bm{x}}}

\def\biu{{\bm{u}}}
\def\biv{{\bm{v}}}
\def\bOmega{{\bm{\Omega}}}
\def\bSigma{{\bm{\Sigma}}}
\def\biV{{\bm{V}}}

\def\btheta{{\bm{\theta}}}
\def\bdelta{{\bm{\delta}}}

\def\bgamma{{\bm{\gamma}}}
\def\E{{\rm E}}

\def\IF{{\mathbb I}}

\def\N{{\rm N}}

\def\tk{\kern 0.08333em}
\def\tn{\kern -0.08333em}
\def\tkk{\kern 0.04167em}
\def\bzero{{\bm{0}}}
\def\bfI{{\bf I}}

\def\th#1{$#1^{\tk{\rm th}}$}

\def\longto{\longrightarrow}

\DeclareMathOperator{\Tr}{Tr}

\DeclareMathOperator{\var}{Var}
\DeclareMathOperator{\cov}{Cov}

\newcommand{\FO}[1]{}

\begin{document}

\title{Cluster-Robust Inference:\\ A Guide to Empirical
Practice\thanks{We are grateful to the editorial board 
for suggesting that we write this paper. We are also grateful to Silvia 
Gon\c calves, Bo Honor\'{e}, Seojeong Lee, and Allan W\"{u}rtz for 
helpful discussions, and to an Associate Editor and two referees for 
valuable comments. MacKinnon and Webb thank the Social Sciences and 
Humanities Research Council of Canada (SSHRC, grants 435-2016-0871 and 
435-2021-0396) for financial support. Nielsen thanks the Danish National
Research Foundation for financial support through a DNRF Chair grant.
The data and programs used in \Cref{sec:empirical} are available at
\url{http://qed.econ.queensu.ca/pub/faculty/mackinnon/guide/}
}}


\author{James G. MacKinnon\thanks{Corresponding author. Address: 
Department of Economics, 94 University Avenue, Queen's University, 
Kingston, Ontario K7L 3N6, Canada. Email:\ \texttt{mackinno@queensu.ca}.
Tel.\ 613-533-2293. Fax 613-533-6668.}\\Queen's University\\
\texttt{mackinno@queensu.ca}  \and
Morten \O rregaard Nielsen\\Aarhus University\\
\texttt{mon@econ.au.dk} \and
Matthew D. Webb\\Carleton University\\
\texttt{matt.webb@carleton.ca}}

\maketitle
\thispagestyle{empty}

\begin{abstract}
Methods for cluster-robust inference are routinely used in economics
and many other disciplines. However, it is only recently that
theoretical foundations for the use of these methods in many
empirically relevant situations have been developed. In this paper, we
use these theoretical results to provide a guide to empirical
practice. We do not attempt to present a comprehensive survey of the
(very large) literature. Instead, we bridge theory and practice by 
providing a thorough guide on what to do and why, based on recently 
available econometric theory and simulation evidence. To practice what 
we preach, we include an empirical analysis of the effects of the 
minimum wage on labor supply of teenagers using individual data.


\vskip 12pt





\end{abstract}

\clearpage
\setcounter{page}{1}
\onehalfspacing

\section{Introduction}
\label{sec:intro}

Ideally, the observations in a sample would be independent of each
other and would each contribute roughly the same amount of information
about the parameter(s) of interest. From the earliest days of
econometrics, it has been recognized that this ideal situation often
does not apply to time\tkk-series data, because there may be serial
correlation. But it has taken much longer for econometricians to
realize that it generally does not apply to cross\tkk-section data
either. The first important step, following \citet{White_1980}, was to
allow for heteroskedasticity of unknown form, and for quite some time
this was the default in empirical work that used cross\tkk-section
data. More recently, however, it has become common for investigators
to drop the assumption of independence as well as the assumption of
homoskedasticity.


There are many ways in which cross\tkk-section observations might be 
dependent, and sometimes it is possible to model this dependence 
explicitly. For example, there is a large literature on spatial 
econometrics and statistics, in which each observation is associated 
with a point in space, and the correlation between any two observations 
is assumed to depend (usually in a rather simple parametric way) on the 
distance between them. See, among many others, \citet{Anselin_1988},
\citet*{Gelfand_2010}, and \citet{Corrado_2012}. However, there are a
great many cases in which either the ``distance'' between any pair of
observations cannot be measured, or the correlation between them is
not related to distance in any way that can readily be modeled.


A more widely applicable approach, on which we focus in this paper, is
to employ cluster-robust inference in the context of least-squares 
estimation. This approach has become increasingly popular over the past 
quarter century and is now used routinely in a great deal of empirical 
microeconomic work. The idea is to divide the sample into $G$ disjoint 
clusters. Depending on the nature of the data, the clusters might 
correspond to classrooms, schools, families, villages, hospitals, firms,
industries, years, cities, counties, states, or countries. This list is 
by no means exhaustive. Any pattern of heteroskedasticity and/or 
dependence is allowed within each cluster, but it is assumed that there 
is independence across clusters and that the assignment of observations
to clusters is known.

Under these assumptions, it is easy to compute cluster-robust standard
errors that can be used to produce asymptotically valid inferences;
see \Cref{sec:CRVE}. However, these inferences may not be at all
reliable in finite samples. Hypothesis tests may reject far more often
than they should. Less commonly, they may reject far less often. In
consequence, the actual coverage of confidence intervals may differ
greatly from their nominal coverage. Therefore, in practice, using
cluster-robust inference often requires a good deal of care.

There are several recent survey papers on cluster-robust inference,
including \citet{CM_2015}, \citet{JGM-CJE}, \citet{Esarey_2019}, and
\citet{MW-survey}. \citet*{CGH_2018} surveys a broader class of
methods for various types of dependent data. Although there will
inevitably be some overlap with these papers, our aim is to provide a
guide to empirical practice rather than a survey of the extant 
literature. We therefore apologize for any missing references and
refer the reader to the survey papers just mentioned for more complete
bibliographies. Our guide is closely based on the econometric theory
and simulation evidence that is currently available. When the theory
is clear and the evidence is strong, we make definitive
recommendations for empirical practice. However, when the theory is
less clear or the evidence is weak, our recommendations are more
guarded.


This guide does not discuss models with clustered data estimated by 
instrumental variables (IV). For such models, neither the current state 
of econometric theory nor the available simulation evidence allows us 
to make recommendations with any confidence. The number of 
over-identifying restrictions and the strength of the instruments can 
greatly affect the reliability of finite\tkk-sample IV inference, and 
dealing with these issues may often be even more important than dealing 
with the issues associated with clustering. There is an enormous 
literature on the topic of weak instruments; see \citet*{Andrews_2019} 
for a recent survey. That paper suggests that, when the disturbances 
of a regression model are independent and homoskedastic, it is generally
possible to obtain reliable (although perhaps imprecise) inferences even
when the instruments are quite weak. However, it also states that this 
is not the case, in general, when there is heteroskedasticity and/or 
clustering.



In \Cref{sec:CRVE}, we obtain the (true) variance matrix for the
coefficient estimators in a linear regression model with clustered
data. The form of this matrix depends on critical assumptions about the
score vectors for each cluster. In practice, inference must be based
on a cluster-robust variance estimator, or CRVE, which estimates the
unknown variance matrix. We discuss the three CRVEs that are commonly
encountered.

\Cref{sec:why} deals with the important and sometimes controversial
issue of when to use cluster-robust inference. It also illustrates how
complicated patterns of intra-cluster correlation can arise in the
context of a simple factor model, introduces the concept of leverage
at the cluster level, discusses the role of cluster fixed effects, and
describes several procedures for deciding the level at which to
cluster.

\Cref{sec:asytheory} concerns the key issue of asymptotic inference.
It explains how to obtain asymptotically valid inferences and
discusses what determines how reliable, or unreliable, these
inferences are likely to be in practice. In many cases, bootstrap
inference tends to be more reliable than asymptotic inference.
\Cref{sec:bootstrap} describes two methods for bootstrap inference,
namely, the pairs cluster bootstrap and the restricted version of the
wild cluster bootstrap, which is called the WCR bootstrap. The former
has the advantage of being applicable to a wide variety of econometric
models, while the latter is only applicable to regression models with
clustered data, for which it typically performs better. For clustered
linear regression models, both of these methods can be remarkably
inexpensive to implement, even for very large samples. We recommend
that, in most cases, the WCR bootstrap be among the methods employed
for inference.

\Cref{sec:other} goes on to discuss some related inferential
procedures. The first of these uses an alternative critical value
estimated from the data, and the second is randomization inference,
which can work well in certain cases where even the WCR bootstrap
fails. \Cref{sec:report} discusses what an empirical investigator
should report in order to convince the reader that results are
reliable. \Cref{sec:empirical} presents an empirical example that uses
individual data to study the effects of the minimum wage on the labor
supply of teenagers. \Cref{sec:conc} provides a summary of the main
points of the paper. This is presented in the form of a short
checklist or guide for empirical researchers on what to do in
practice, with references to relevant sections.


\section{Cluster-Robust Variance Estimators}
\label{sec:CRVE}


\subsection{The Clustered Regression Model}
\label{subsec:Model}



Throughout the paper, we deal with the linear regression model $y_i =
\bix_i^\top\tn\bbeta + u_i$, which, if the data have been divided into
$G$ disjoint clusters, can be rewritten as
\begin{equation}
\label{eq:lrmodel} 
\biy_g =\biX_g\bbeta + \biu_g, \quad g=1,\ldots,G.
\end{equation}
Here $\biX_g$ is an $N_g\times k$ matrix of exogenous regressors, 
$\bbeta$ is a $k$-vector of coefficients, $\biy_g$ is an $N_g$-vector
of observations on the regressand, and $\biu_g$ is an $N_g$-vector of
disturbances (or error terms). Thus $\biX_g$, $\biy_g$, and $\biu_g$
stack the $\bix_i^\top$, $y_i$, and $u_i$, respectively. In many
cases, the regressors will include cluster fixed effects; see
\Cref{subsec:FE}.  Since the \th{g} cluster has $N_g$ observations,
the sample size is $N = \sum_{g=1}^G N_g$. The $\biX_g$ may of course
be stacked into an $N\times k$ matrix $\biX$\tn, and likewise the
$\biy_g$ and $\biu_g$ may be stacked into $N$-vectors $\biy$ and
$\biu$, so that \eqref{eq:lrmodel} can be rewritten in the usual way
as $\biy = \biX\!\bbeta + \biu$.

It is assumed that the data are generated by \eqref{eq:lrmodel} with
$\bbeta=\bbeta_0$. Under this assumption, the OLS estimator of
$\bbeta$ is
\begin{equation*}
\hat\bbeta = (\biX^\top\biX)^{-1}\biX^\top\biy
= \bbeta_0 +  (\biX^\top\biX)^{-1}\biX^\top\biu.
\end{equation*}
It follows that
\begin{equation}
\hat\bbeta - \bbeta_0 = (\biX^\top\biX)^{-1}\sum_{g=1}^G \biX_g^\top\biu_g
= \Big(\tn\sum_{g=1}^G\biX_g^\top\biX_g\Big)^{\!-1} \sum_{g=1}^G \bis_g,
\label{eq:betahat}
\end{equation}
where $\bis_g = \biX_g^\top \biu_g$ denotes the $k\times1$ score
vector corresponding to the \th{g} cluster. For a correctly specified 
model, $\E(\bis_g)=\bzero$ for all $g$. From the rightmost expression
in \eqref{eq:betahat}, the distribution of the OLS estimator
$\hat\bbeta$ depends on $\biu$ only through the distribution of the
score vectors $\bis_g$. Ideally, the sum of the $\bis_g$, suitably
normalized, would be well approximated by a multivariate normal
distribution with mean zero.


Because we can always divide the sample into $G$ clusters in any way
we like, \eqref{eq:betahat} is true for any distribution of the
disturbance vector $\biu$. Dividing the sample into clusters only
becomes meaningful if we further assume that
\begin{equation}
\label{eq:Sigma_g}
\E(\bis_g\bis_g^\top) = \bSigma_g \quad\mbox{and}\quad
\E(\bis_g\bis_{g'}^\top) = \bzero, \quad g,g'=1,\ldots,G,\quad g'\ne g,
\end{equation}
where the variance matrix of the scores for the \th{g} cluster,
$\bSigma_g$, is a $k\times k$ symmetric, positive semidefinite matrix.
The second assumption in \eqref{eq:Sigma_g} is the key one. It states
that the scores for every cluster are uncorrelated with the scores for
every other cluster. In contrast, the first assumption imposes no real
limitations, so that the $\bSigma_g$ matrices may display any patterns
of heteroskedasticity and/or within-cluster dependence. Indeed, one
motivation for using cluster-robust inference is that it is robust
against both heteroskedasticity and intra-cluster dependence without
imposing any restrictions on the (unknown) form of either of them.



For now, we will simply assume that \eqref{eq:Sigma_g} holds for some
specified division of the observations into clusters. Although the 
choice of clustering structure is often controversial, or at least 
somewhat debatable, the structure is almost always assumed known in 
both theoretical and applied work. The important issue of how to choose 
the clustering structure will be discussed below in \Cref{subsec:level}.


It follows immediately from \eqref{eq:betahat} that an estimator of the 
variance of $\hat\bbeta$ should be based on the usual sandwich formula,
\begin{equation}
\label{eq:trueV}
(\biX^\top\biX)^{-1} \Big(\tn\sum_{g=1}^G \bSigma_g\tn\Big) 
(\biX^\top\biX)^{-1}.
\end{equation}
Of course, this matrix cannot be computed, because we need to estimate
the $\bSigma_g$. This can be done in several ways, as we discuss in
\Cref{subsec:CVmats}.


As \eqref{eq:betahat} makes clear, it is the properties of the score
vectors that matter for inference. Of course, those properties are
inherited from the properties of the disturbances and the regressors.
If $\bOmega_g = \E (\biu_g\biu_g^\top |\biX)$ denotes the conditional
variance matrix of $\biu_g$, then
\begin{equation}
\bSigma_g = \E(\biX_g^\top\bOmega_g\biX_g), \quad g=1,\ldots,G.
\label{sigomega}
\end{equation}
Thus, instead of making assumptions directly about the $\bSigma_g$, as
we did in \eqref{eq:Sigma_g}, it may be more illuminating to make
assumptions about the $\bOmega_g$ and the $\biX_g$. If
$\E(\biu_g\biu_{g'}^\top |\biX) = \bzero$ for all $g' \neq g$, then
the second assumption in \eqref{eq:Sigma_g} will hold. It will also
hold if the regressors are exogenous and uncorrelated across clusters
even when the disturbances are not. 

Since the score vector $\bis_g$ can be written as $\sum_{i=1}^{N_g}
\bis_{gi} = \sum_{i=1}^{N_g} \biX^\top_{gi} u_{gi}$, where $\biX_{gi}$
is the \th{i} row of $\biX_g$ and $u_{gi}$ is the \th{i} element of
$\biu_g$, the outer product of the score vector with itself is seen to
be
\begin{equation}
\bis_g\bis_g^\top = 
\Big(\tn\sum_{i=1}^{N_g} \biX^\top_{gi} u_{gi}\Big)
\Big(\tn\sum_{i=1}^{N_g} \biX^\top_{gi} u_{gi}\Big)^{\!\!\top} =
\sum_{i=1}^{N_g} \sum_{j=1}^{N_g} \biX^\top_{gi} \biX_{gj} u_{gi} u_{gj}
= \sum_{i=1}^{N_g} \sum_{j=1}^{N_g} \bis_{gi} \bis_{gj}^\top.
\label{eq:opscores}
\end{equation}
When $\E(u_{gi}^2 | \biX) = \sigma^2$ and $\E(u_{gi} u_{gj} | \biX) =
0$ for $i \neq j$, then $\E ( \bis_g\bis_g^\top | \biX ) = \sigma^2 
\biX_g^\top\tn \biX_g$. In that case, we would replace $\bSigma_g$
with $\sigma^2 (\biX_g^\top\biX_g )$ in \eqref{eq:trueV} and obtain
the classic result that $\var (\hat\bbeta | \biX ) =
\sigma^2(\biX^\top \biX)^{-1}$.



Taking expectations in \eqref{eq:opscores} and defining the covariance
matrix $\bSigma_{g,ij} = \E ( \bis_{gi}\bis^\top_{gj} )$, we find
that, in general, $\bSigma_g = \sum_{i=1}^{N_g}\sum_{j=1}^{N_g}\E
(\bis_{gi} \bis_{gj}^\top)
=\sum_{i=1}^{N_g}\sum_{j=1}^{N_g}\bSigma_{g,ij}$. In the special case
where the score vectors $\bis_{gi}$ are uncorrelated within each
cluster, i.e.\ where $\bSigma_{g,ij}=\bzero$ for $i \neq j$, we find
that $\bSigma_g = \sum_{i=1}^{N_g}\E (\bis_{gi}\bis_{gi}^\top) =
\sum_{i=1}^{N_g} \bSigma_{g,ii}$. The difference between these two 
expressions for $\bSigma_g$ is
\begin{equation}
\sum_{i=1}^{N_g} \sum_{j=1}^{N_g} \E\big(\bis_{gi}\bis_{gj}^\top\big)
- \sum_{i=1}^{N_g} \E\big(\bis_{gi} \bis_{gi}^\top\big) 
= \sum_{i=1}^{N_g} \sum_{j\ne i} \bSigma_{g,ij}.
\label{eq:diffop}
\end{equation}
The rightmost expression in \eqref{eq:diffop} is just the summation of
the $N_g^2 - N_g$ matrices that correspond to the off-diagonal
elements of $\bSigma_g$. It equals zero whenever there is no
intra-cluster correlation, but in general it is $O(N_g^2)$. Therefore,
incorrectly assuming that the scores are not correlated within
clusters potentially leads to much larger errors of inference when
clusters are large than when they are small. For sufficiently large
values of $N_g$, these errors may be large even when all of the
$\bSigma_{g,ij}$ for $i\ne j$ are very small \citep{JGM_2016}.

The famous ``Moulton factor'' \citep{Moulton_1986} gives the ratio of
the true variance of an OLS coefficient, from \eqref{eq:trueV}, to the
variance based on the classic formula $\sigma^2(\biX^\top\biX)^{-1}$
under the assumption that both the disturbances and the regressor of 
interest (after other regressors have been partialed out) are 
equi-correlated within clusters; see \Cref{subsec:sources}. If the
scores were scalars with intra-cluster correlation $\rho_s$, and the
cluster sizes were constant, say $N_g = M$\tn, then the Moulton factor
would be $1+(M-1)\rho_s$. The second term is proportional to the
number of observations per cluster, so the mistakes made by not
clustering can be enormous when clusters are large.

Since the disturbances in \eqref{eq:lrmodel} are neither independent
nor homoskedastic, it seems relevant to consider GLS estimation, even
though OLS estimation is almost always used in practice. If we were
willing to specify a simple parametric form for the $\bOmega_g$
matrices, then we could use feasible GLS. For example, if we assumed
that the disturbances were equi-correlated within each cluster, that
would be equivalent to specifying a random-effects model; see
\Cref{subsec:sources}. In practice, however, the regressors in
\eqref{eq:lrmodel} very often include cluster fixed effects
(\Cref{subsec:FE}), and the latter remove whatever intra-cluster
correlation a random-effects specification induces. So we would need
to specify a more complicated model if we wanted to use feasible GLS.
In any case, specifying a parametric form for the intra-cluster
correlations would imply making assumptions much stronger than those
in \eqref{eq:Sigma_g}, and this would violate the principal objective
of cluster-robust inference, namely, to be robust to arbitrary and
unknown dependence and heteroskedasticity within clusters.

\subsection{Three Feasible CRVEs}
\label{subsec:CVmats}

The natural way to estimate \eqref{eq:trueV} is to replace the
$\bSigma_g$ matrices by their empirical counterparts, which are the
outer products of the empirical score vectors $\hat\bis_g =
\biX_g^\top \hat\biu_g$ with themselves.  If, in addition, we multiply
by a correction for degrees of freedom, we obtain
\begin{equation}
\mbox{CV$_{\tn1}$:}\qquad
\frac{G(N-1)}{(G-1)(N-k)}
(\biX^\top\biX)^{-1}
\Big(\tn\sum_{g=1}^G \hat\bis_g\hat\bis_g^\top\Big) (\biX^\top\biX)^{-1}.
\label{eq:CV1}
\end{equation}
At present, this is by far the most widely used CRVE in practice.
Observe that, when $G=N$\tn, CV$_{\tn1}$ reduces to the familiar
HC$_1$ estimator \citep{MW_1985} that is robust only to
heteroskedasticity of unknown form.


The empirical score vectors $\hat\bis_g$ are not always good
estimators of the $\bis_g$. CV$_{\tn1}$ attempts to compensate for
this by including a degrees-of-freedom factor. Two alternative CRVEs,
proposed in \citet{BM_2002}, instead replace the empirical score
vectors $\hat\bis_g$ by modified score vectors that use transformed
residuals. The first of these is
\begin{equation}
\mbox{CV$_{\tn2}$:}\qquad
(\biX^\top\biX)^{-1}\Big(\tk\sum_{g=1}^G \grave\bis_g\grave\bis_g^\top
\Big)(\biX^\top\biX)^{-1},
\label{eq:CV2}
\end{equation}
where $\grave\bis_g = \biX_g^\top\biM_{gg}^{-1/2}\tk\hat\biu_g$, with
$\biM_{gg} = \bfI_{N_g} - \biX_g(\biX^\top\biX)^{-1}\tn\biX_g^\top$.
Thus $\biM_{gg}$ is the \th{g} diagonal block of the projection matrix
$\biM_\biX$, which satisfies $\hat\biu = \biM_\biX\biu$, and
$\biM_{gg}^{-1/2}$ is its inverse symmetric square root. The CV$_2$ 
estimator reduces to the familiar HC$_2$ estimator when $G=N$. If the
variance matrix of every $\biu_g$ were proportional to an identity
matrix, then CV$_{\tn2}$ would actually be unbiased \citep{PT_2018}.




The second alternative CRVE is
\begin{equation}
\mbox{CV$_{\tn3}$:}\qquad
\frac{G-1}{G}
(\biX^\top\biX)^{-1}\Big(\tk\sum_{g=1}^G \acute\bis_g\acute\bis_g^\top
\Big)(\biX^\top\biX)^{-1},
\label{eq:CV3}
\end{equation}
where $\acute\bis_g = \biX_g^\top\biM_{gg}^{-1}\tk\hat\biu_g$. As we
discuss in \Cref{subsec:leverage}, CV$_{\tn3}$ is actually a jackknife
estimator which generalizes the familiar HC$_3$ estimator of
\citet{MW_1985}.

As written in \eqref{eq:CV2} and \eqref{eq:CV3}, both CV$_{\tn2}$ and
CV$_{\tn3}$ are computationally infeasible for large samples, because
they involve the $N_g\times N_g$ matrices $\biM_{gg}$. However,
\citet*{NAAMW_2020} proposes a more efficient algorithm for both of
them, and \citet*{MNW-influence} provides an even more efficient one
for CV$_{\tn3}$ by exploiting the fact that it is a jackknife
estimator; see \Cref{subsec:leverage}. Moreover, when one of the
regressors is a fixed-effect dummy for cluster $g$, the $\biM_{gg}$ 
matrices are singular. This problem can be avoided, and some computer
time saved, by partialing out the fixed-effect dummies as discussed in
\Cref{subsec:leverage} and \citet{PT_2018}.

%




It seems plausible that both CV$_{\tn2}$ and CV$_{\tn3}$ should
perform better in finite samples than CV$_{\tn1}$, because the
modified score vectors $\grave\bis_g$ and $\acute\bis_g$ ought to
provide better approximations to the $\bis_g$ than do
the~$\hat\bis_g$. We would expect tests based on CV$_{\tn3}$ to be
more conservative than ones based on CV$_{\tn2}$, just as ones based
on HC$_3$ are more conservative than ones based on HC$_2$, because the
$\acute\bis_g$ are ``shrunk'' more than the $\grave\bis_g$. Simulation
evidence dating back to \citet{BM_2002} suggests that CV$_{\tn3}$
typically yields the most reliable tests, but that they can sometimes
under-reject; see also \citet*{MNW-bootknife}.


\section{Why and How to Cluster}
\label{sec:why}

We cannot hope to obtain reliable inferences when using clustered data
unless we know the actual clustering structure, at least to a good
approximation. Thus, before specifying any clustering structure, we
need to think about how intra-cluster correlations may arise and why
independence across clusters may, or may not, be plausible for that
structure.


\citet*{AAIW_2017} distinguishes between two alternative approaches to
inference, referred to as ``model-based'' and ``design-based.'' The
model-based approach is the traditional one, according to which every
sample is treated as a random outcome, or realization, from some
data-generating process (DGP), which in our context is the model 
\eqref{eq:lrmodel}, and the objective is to draw inferences about 
parameters in the DGP, which are interpreted as features of the 
population. In particular, the DGP is the source of randomness and 
hence an important determinant of the clustering structure.


The ``design-based'' approach to inference, which is analyzed in
detail in \citet{AAIW_2017}, is conceptually different from the
model-based framework. It involves thinking about the population,
estimand, sampling scheme, parameters of interest, and even the notion
of statistical uncertainty in a different manner. For example, according
to the design-based approach, statistical uncertainty does not derive 
from a DGP, but rather is induced solely by the sampling uncertainty 
coming from sampling from a fixed population \citep{AAIW_2020}. 
Therefore, under this approach, the sampling process is an important 
determinant of the clustering structure.


In some cases, the two approaches make use of similar inferential
procedures, but in others they employ quite different ones. Both
approaches may be informative about the choice of whether to cluster
and at which level, although their motivations may differ. We follow 
most of the existing literature and focus exclusively on the model-based
approach in the remainder of this paper.

\subsection{Modeling Intra-Cluster Dependence}
\label{subsec:sources}


Intra-cluster correlations of the disturbances and regressors, and
hence of the scores, can arise for many reasons. By making the
assumptions in \eqref{eq:Sigma_g} and using cluster-robust inference,
we avoid the need to model these correlations. Nevertheless, it can be
illuminating to consider such models in order to learn about 
intra-cluster dependence and their consequences. The simplest and most
popular model is the random-effects, or error-components, model
\begin{equation}
\label{eq:REmodel}
u_{gi} = \lambda\tkk \varepsilon_g + \varepsilon_{gi},
\end{equation}
where $u_{gi}$ is the disturbance for observation $i$ within cluster
$g$, $\varepsilon_{gi} \sim {\rm iid}(0, \omega^2)$ is an
idiosyncratic shock for observation $i$, $\varepsilon_g \sim {\rm
iid}(0, 1)$ is a cluster-wide shock for cluster $g$, and the two
shocks are independent. The model \eqref{eq:REmodel} implies that the
variance matrix $\bOmega_g$ of the $u_{gi}$ for cluster $g$ has a very
simple form with diagonal elements equal to $\lambda^2 + \omega^2$
and off-diagonal elements equal to $\lambda^2$. Thus the disturbances 
within every cluster are equi-correlated, with correlation coefficient 
$\lambda^2/(\lambda^2 + \omega^2)$.

Although the random-effects model \eqref{eq:REmodel} has a long and
distinguished history, it is almost certainly too simple. As we
discuss in \Cref{subsec:FE}, it is very common in modern empirical
practice to include a set of cluster fixed effects among the regressors 
in \eqref{eq:lrmodel}. In the case of \eqref{eq:REmodel}, these fixed 
effects are simply estimates of the $\lambda\tkk \varepsilon_g$, and by
including them we therefore remove all of the intra-cluster correlation.
Thus, if the random-effects model \eqref{eq:REmodel} were correct, 
there would be no need to worry about cluster-robust inference whenever 
the regressors included cluster fixed effects. Note also that inclusion
of cluster fixed effects usually comes at the price of larger standard 
errors on the coefficients of interest.



In practice, however, it usually seems to be the case that we need
both cluster fixed effects and a CRVE; see
\Cref{subsec:FE,subsec:placebo}. This implies that whatever process is
generating the intra-cluster correlations must be more complicated
than \eqref{eq:REmodel}. A simple example is the (very standard)
factor model 
\begin{equation}
\label{eq:factor}
u_{gi} = \lambda_{gi}\tk \varepsilon_g + \varepsilon_{gi},
\end{equation}
which differs from \eqref{eq:REmodel} in one important respect. The
effect of the cluster-wide shock $\varepsilon_g$ on $u_{gi}$ is given
not by a coefficient $\lambda$ but by a weight, or factor loading,
$\lambda_{gi}$. These factor loadings could be either fixed parameters
or random variables. They determine the extent to which observation
$i$ within cluster $g$ is affected by the cluster-wide shock
$\varepsilon_g$.

As an example, if the observations were for individual students, the
clusters denoted classrooms, and the outcome were student achievement,
then $\varepsilon_{gi}$ would measure unobserved student-specific
characteristics, $\varepsilon_g$ would measure unobserved teacher
quality (and perhaps other features of the class), and $\lambda_{gi}$
would measure the extent to which the disturbance term for student $i$
is affected by teacher quality. Clearly, the $\lambda_{gi}$ do not
need to be the same for all~$i$. Similar motivating examples based on
\eqref{eq:factor} can easily be given in many fields, including labor
economics, health economics, development economics, and financial
economics.

To verify that the factor model in \eqref{eq:factor} generates 
dependence within clusters, it suffices to derive the second-order
moments of the $u_{gi}$. We find that $\E (u_{gi}) = 0$ and $\var
(u_{gi}) = \lambda_{gi}^2 + \omega^2$. The cluster dependence is
characterized by $\cov (u_{gi},u_{gj}) = \lambda_{gi} \lambda_{gj}$,
which differs across $(i,j)$ pairs and is zero only when the factor
loadings are zero. In the context of the classroom example, the
intra-cluster covariances would be zero only if the teacher had no
effect on student achievement. Moreover, the correlations would be
fully captured by classroom fixed effects if and only if
$\lambda_{gi}$ were the same for all~$i$.

The factor model in \eqref{eq:factor} is discussed in terms of the 
disturbances $u_{gi}$ rather than the scores. There are at least two
simple cases in which the same model structure, and in particular the
same within-cluster correlation structure, applies to the scores. The
first is when a regressor is generated by a model similar to
\eqref{eq:factor}, but possibly with different parameters. The second
is when a regressor only varies at the cluster level, as is often the
case for dummy variables, especially treatment dummies.

The model \eqref{eq:factor} has only one clustering dimension, but the
idea does not apply exclusively to cross\tkk-section data. For
example, if the observations also had a time dimension, we could
replace each of the $\varepsilon_g$ by a time\tkk-series process at
the cluster level. This would yield a pattern where, within a cluster,
observations that were closer together in time would be more
correlated than observations that were further apart. In
\Cref{subsec:emp-placebo}, we generate placebo regressors in this way.
For panel data, it is possible that, in addition to correlation within
cross\tkk-sectional units across time periods, there may be
correlation within time periods across cross\tkk-sectional units. This
leads to two\tkk-way clustering, which is discussed in
\Cref{subsec:twoway}.


In principle, we might be able to estimate a factor model like
\eqref{eq:factor} and use it to obtain feasible GLS estimates, as
mentioned at the end of \Cref{subsec:Model}. However, this would be
relatively complicated and rather arbitrary, since there are many
plausible ways in which the details of \eqref{eq:factor} could be
specified. Moreover, any factor model would necessarily impose far
stronger restrictions than the weak assumptions given in
\eqref{eq:Sigma_g}. Thus estimating any sort of factor model would
inevitably require far more effort, and surely result in much more
fragile inferences, than simply employing OLS estimation together with
a~CRVE.

\subsection{Do Cluster Fixed Effects Remove Intra-Cluster Dependence?}
\label{subsec:FE}




Investigators very often include fixed effects at the cluster level
among the regressors. There are generally good reasons for doing so.
The cluster fixed effects implicitly model a large number of possibly
omitted explanatory variables without assuming, implausibly, that the
omitted variables are uncorrelated with the included ones.

It is sometimes believed that fixed effects remove any within-cluster
dependence and hence eliminate the need to use a CRVE. However, as was
pointed out by \citet{Arellano_1987}, that is in fact only true under
very special circumstances. Including cluster fixed effects in any
regression model forces the intra-cluster sample average to be zero
for each cluster. In particular, including cluster fixed effects
transforms the factor model \eqref{eq:factor} into
\begin{equation}
\label{eq:FEscore}
u_{gi} - \bar{u}_g = ( \lambda_{gi} -\bar\lambda_g ) 
\varepsilon_g + ( \varepsilon_{gi} - \bar\varepsilon_g ),
\end{equation}
where the averages are taken across observations within each cluster, 
so that, for example, $\bar{u}_g = N_g^{-1}\sum_{i=1}^{N_g}u_{gi}$. The 
intra-cluster covariance for \eqref{eq:FEscore} is
\begin{equation}
\label{eq:FEcov}
\cov (u_{gi}- \bar{u}_g, u_{gj}-\bar{u}_g) =
( \lambda_{gi} -\bar\lambda_g )( \lambda_{gj}-\bar\lambda_g ),
\end{equation}
which is zero if and only if $\lambda_{gi}$ is the same for all~$i$.
In other words, the random-effects model \eqref{eq:REmodel} is the
\emph{only} model within the class of factor models \eqref{eq:factor}
for which including cluster fixed effects can remove all intra-cluster
dependence. Some dependence necessarily remains whenever there is any
variation in factor loadings across observations within clusters.

Furthermore, \eqref{eq:FEcov} strongly suggests that, whether or not a
regression model includes cluster fixed effects, the scores will tend
to be clustered whenever within-cluster dependence can be approximated
by a factor model like \eqref{eq:factor}. Including fixed effects will
almost always reduce the intra-cluster correlations, but rarely will it 
entirely eliminate them. Because even very small intra-cluster 
correlations can have a large effect on standard errors when the 
clusters are large (see \eqref{eq:diffop} and the discussion that 
follows) it generally seems unwise to assume that cluster fixed effects 
make it unnecessary to use a~CRVE.

In view of these arguments, it has become quite standard in modern
empirical practice both to include cluster fixed effects (and perhaps
other fixed effects as well) and also to employ cluster-robust
inference. The empirical example in \Cref{sec:empirical} is typical in
these respects. Of course, cluster fixed effects cannot be included
when the regressor of interest is a treatment dummy and treatment is
at the cluster level, since the treatment dummy and the fixed effects
would be perfectly collinear. This problem does not arise for
difference\tkk-in-differences (DiD) regressions,
because only some observations in the treated clusters are treated. In
recent empirical work with non-staggered adoption of treatment, the
regressions almost always include at least two sets of fixed effects,
one for time periods and one for cross-sectional units, with
clustering typically by the latter; see \Cref{subsec:placebo}.




\subsection{At What Level Should We Cluster?}
\label{subsec:level}

In many cases, there is more than one level at which we could cluster.
For example, with data on educational outcomes, we may be able to
cluster by classroom, by school, or perhaps by school district. With
data that are coded geographically, we may be able to cluster by
county, by state, or even by region. Choosing the right level at which
to cluster is not always easy, and choosing the wrong level can have
serious consequences.

Suppose, for concreteness, that there are two possible levels of
clustering, coarse and fine, with one or more fine clusters nested
within each of the coarse clusters. When there are $G$ coarse
clusters, the middle matrix in \eqref{eq:trueV} is $\sum_{g=1}^G
\bSigma_g$. If each coarse cluster contains $M_g$ fine clusters
indexed by $h$, then $\bSigma_g$ can be written as
\begin{equation}
\label{eq:midcoarse}
\bSigma_g = \sum_{h_1=1}^{M_g} \sum_{h_2=1}^{M_g} \bSigma_{g,h_1h_2},
\end{equation}
where $\bSigma_{g,h_1h_2}$ denotes the covariance matrix of the scores 
for fine clusters $h_1$ and $h_2$ within coarse cluster~$g$. Under the 
assumption of fine clustering, $\bSigma_{g,h_1h_2} = \bSigma_{gh}$ 
when $h_1=h_2=h$ and $\bSigma_{g,h_1h_2} = \bzero$ when $h_1\ne h_2$, 
so that the middle matrix in \eqref{eq:trueV} reduces to 
$\sum_{g=1}^G \sum_{h=1}^{M_g} \bSigma_{gh}$.

From \eqref{eq:midcoarse}, the difference between the middle matrices
for coarse and fine clustering is
\begin{equation}
\label{eq:middiff}
\sum_{g=1}^G \bSigma_g - \sum_{g=1}^G \sum_{h=1}^{M_g} \bSigma_{gh} =
2 \sum_{g=1}^G \, \sum_{h_1=1}^{M_g}\,
\sum_{h_2=h_1+1}^{M_g}\!\!\!\bSigma_{g,h_1h_2}.
\end{equation}
The finest possible level of clustering is no clustering at all. In
that case, the right-hand side of \eqref{eq:middiff} reduces to the
right-hand side of \eqref{eq:diffop}, because the fine clusters within
each coarse cluster are just the individual observations.


Under the assumption of fine clustering, the terms on the right-hand
side of \eqref{eq:middiff} are all equal to zero. Under the assumption
of coarse clustering, however, at least some of them are non-zero, and
\eqref{eq:middiff} must therefore be estimated. If we cluster at the
fine level when coarse clustering is appropriate, the CRVE is
inconsistent. On the other hand, if we cluster at the coarse level
when fine clustering is appropriate, the CRVE has to estimate
\eqref{eq:middiff} even though it is actually zero. This makes the
CRVE less efficient than it should be, leading to loss of power, or,
equivalently, to confidence intervals that are unnecessarily long,
especially when the number of coarse clusters is small. 


Using simulation methods, \citet{MW-survey} investigates the
consequences on hypothesis tests of clustering at an incorrect level. 
Clustering at too fine a level generally leads to serious 
over-rejection, which becomes worse as the sample size increases with 
the numbers of clusters at all levels held constant. This is exactly 
what we would expect; see the discussion following \eqref{eq:diffop}. 
Clustering at too coarse a level also leads to both some over-rejection 
and some loss of power, especially when the number of clusters is small.

Two rules of thumb are commonly suggested for choosing the right
level of clustering. The simplest is just to cluster at the coarsest
feasible level \citep[Section~IV]{CM_2015}. This may be attractive
when the number of coarse clusters $G$ is reasonably large, but it can
be dangerous when $G$ is small, or when the clusters are heterogeneous
in size or other features; see \Cref{subsec:failure}.

A more conservative rule of thumb is to cluster at whatever level
yields the largest standard error(s) for the coefficient(s) of
interest \citep[Section~8.2]{MHE_2008}. This rule will often lead to
the same outcome as the first one, but not always. When $G$ is small,
cluster-robust standard errors tend to be too small, sometimes much
too small (\Cref{subsec:failure}). Hence, the second rule of thumb is
considerably less likely to lead to severe over-rejection than the
first one. However, because it is conservative, it can lead to loss of
power (or, equivalently, confidence intervals that are unnecessarily
long).

When the regressor of interest is a treatment dummy, and the level at
which treatment is assigned is known, then it generally makes sense
to cluster at that level \citep*{BDM_2004}. If treatment is assigned by
cluster, whether for all observations in each cluster or just for some
of them, as in the case of DiD models, then the scores will be
correlated within the treated clusters whenever there is any
intra-cluster correlation of the disturbances. Thus it never makes
sense to cluster at a level finer than the one at which treatment is
assigned. If we are certain that clusters are treated at random, then
it also does not make sense to cluster at a coarser level. However,
when there are two or more possible levels of clustering, it may not
be realistic to assume that treatments are independent across
finer-level clusters within the coarser-level ones. Unless we are
certain that this is actually the case, it may be safer to cluster at
a coarser level than the one at which treatment was supposedly
assigned. For example, it may make sense to cluster by school instead
of by classroom even when treatment was supposed to be assigned by
classroom.


Instead of using a rule of thumb, we can test for the correct level of
clustering. The best-known such test is an ingenious but indirect one
proposed in \citet{IM_2016}. It requires the model to be estimated
separately for every coarse cluster, something that is not possible
when the regressor of interest is invariant within some clusters, as
is typically the case for treatment models and DiD models. It is also
invalid if the parameter of interest has different meanings for 
different clusters; see \Cref{subsec:small}. When it is valid, the
test statistic compares the observed variation of the estimates across
clusters with an estimate of what that variation would be if
clustering were actually at a finer level.

\citet*{MNW-testing} proposes direct tests called score\tkk-variance
tests, which compare the variance of the scores for two nested levels
of clustering. For example, when there is just one coefficient, the
empirical analog of \eqref{eq:middiff} is a scalar that, when divided
by the square root of an estimate of its variance, is asymptotically
distributed as~$\N(0,1)$. The null hypothesis is that the (true)
standard errors are the same for fine and coarse clustering, and the
(one\tkk-sided) alternative is that they are larger for the latter
than for the former. \citet{MNW-testing} also proposes wild (cluster)
bootstrap implementations of these tests to improve their
finite\tkk-sample properties.

\citet{Cai_2021} takes a different approach, proposing a test for the
level of clustering that is based on randomization inference
(\Cref{subsec:RI}). This test is designed for settings with a small
number of coarse clusters and a small number of fine clusters within
each of them.

It seems natural to cluster at the coarse level when a test rejects
the null hypothesis, and to cluster at the fine level when it does
not.  However, choosing the level of clustering in this way is a form
of pre\tkk-testing, which can lead to estimators with distributions
that are poorly approximated by asymptotic theory, even in large
samples \citep{LP_2005}. Using a pre\tkk-test in this way will 
inevitably lead to over-rejection when there is actually coarse 
clustering, the standard errors for coarse clustering are larger than 
the ones for fine clustering, and the test incorrectly fails to reject 
the null hypothesis. Thus using such a test is less conservative than 
relying on the second rule of thumb discussed above. On the other hand, 
we may feel more comfortable with the second rule of thumb when it 
agrees with the outcomes of one or more tests for the level of 
clustering.

\subsection{Leverage and Influence}
\label{subsec:leverage}


As will be explained in \Cref{sec:asytheory}, asymptotic inference
depends on being able to apply laws of large numbers and central limit
theorems to functions of the (empirical) score vectors. How well those
theorems work depends on how homogeneous the score vectors are across
clusters. When they are quite heterogeneous, asymptotic inference may
be problematic; see \Cref{subsec:failure}. It is therefore desirable
to measure the extent of cluster-level heterogeneity.

Classic measures of observation-level heterogeneity are leverage and 
influence \citep*{BKW_1980,CH_1986}. These are generalized to 
cluster-level measures in \citet{MNW-influence}. One possible
consequence of heterogeneity is that the estimates may change a lot
when certain clusters are deleted. When this is the case, a cluster is
said to be influential. In order to identify individually influential
clusters, first construct the matrices $\biX_g^\top\biX_g$ and the
vectors $\biX_g^\top\biy_g$, for $g=1,\ldots,G$. Then
\begin{equation}
\label{eq:delone}
\hat\bbeta^{(g)} = (\biX^\top\biX - \biX_g^\top\biX_g)^{-1}
(\biX^\top\biy - \biX_g^\top\biy_g)
\end{equation}
is the vector of least squares estimators when cluster $g$ is deleted.
It should not be expensive to compute $\hat\bbeta^{(g)}$ for every
cluster using~\eqref{eq:delone}. Note, however, that we cannot partial
out regressors other than cluster fixed effects (see below) prior to
computing the $\hat\bbeta^{(g)}$, because the latter would then depend
indirectly on the observations for the \th{g} cluster.



When there is a parameter of particular interest, say~$\beta_j$, then
it will often be a good idea to report the $\hat\beta_j^{(g)}$ for 
$g=1,\ldots,G$ in either a histogram or a table. If
$\hat\beta_j^{(h)}$ differs a lot from $\hat\beta_j$ for some cluster
$h$\tn, then cluster $h$ is evidently influential. In a few extreme
cases, there may be a cluster $h$ for which it is impossible to
compute $\hat\beta_j^{(h)}$\tn. If so, then the original estimates
should probably not be believed. This will happen, for example, when
cluster $h$ is the only treated one, and we will see in
\Cref{subsubsec:few} that inference is extremely unreliable in that
case.



The $\hat\bbeta^{(g)}$ are of interest even when there is no reason to
expect any clusters to be influential. As \citet{MNW-bootknife} shows,
an alternative way to write CV$_{\tn3}$ is 
\begin{equation}
\mbox{CV$_{\tn3}$:}\qquad
\frac{G-1}{G} \sum_{g=1}^G (\hat\bbeta^{(g)} - \hat\bbeta)
(\hat\bbeta^{(g)} - \hat\bbeta)^\top.
\label{eq:jackvar}
\end{equation}
This is the matrix version of the classic jackknife variance estimator
given in \citet{Efron_81} and others. Unless all clusters are very
small, \eqref{eq:jackvar} is enormously faster to compute than
\eqref{eq:CV3}. The \texttt{Stata} command \texttt{summclust}, which
is described in detail in \cite{MNW-influence}, calculates CV$_{\tn3}$
standard errors based on \eqref{eq:jackvar}.




As pointed out in \citet{BKW_1980} and \citet{CH_1986}, it is often 
valuable to identify high-leverage observations as well as influential
ones. It is perhaps even more valuable to identify high-leverage
clusters \citep{MNW-influence}. Loosely speaking, a high-leverage
cluster is one whose regressors contain a lot of information. At the
observation level, high-leverage observations are associated with a
high value of $h_i$, the \th{i} diagonal element of $\biH = \biP_\biX
= \biX (\biX^\top\biX)^{-1}\biX^\top$. The analog of $h_i$ in the
cluster case is the $N_g\times N_g$ matrix $\biH_g =
\biX_g(\biX^\top\biX)^{-1} \biX_g^\top$\tn. Since it is not feasible
to report the $\biH_g$, we suggest that investigators instead report
their traces, which are
\begin{equation}
\label{eq:traceXX}
L_g = \Tr(\biH_g) =
\Tr\!\big(\biX_g^\top\!\biX_g(\biX^\top\!\biX)^{-1}\big), \quad
g=1,\ldots,G.
\end{equation}
These are easy to compute because we have already calculated
$(\biX^\top\biX)^{-1}$ and the $\biX_g^\top\biX_g$. For any cluster
that contains just one observation, $L_g$ reduces to the usual measure
of leverage at the observation level. High-leverage clusters can be
identified by comparing the $L_g$ to their own average, which is
$k/G$. If, for some $h$, $L_h$ is substantially larger than $k/G$,
then cluster $h$ has high leverage. This can happen either because
$N_h$ is much larger than $G/N$ or because the matrix $\biX_h$ is
somehow extreme relative to the other $\biX_g$ matrices, or both. For
example, $L_h$ is likely to be much larger than $k/G$ if cluster $h$
is one of just a few treated clusters. 

Regression models often include cluster fixed effects. It is
computationally attractive to partial them out before estimation
begins, using for example the \texttt{areg} procedure in
\texttt{Stata}. When one of the regressors is a fixed-effect dummy for
cluster $g$, the matrices $\biX^\top\biX - \biX_g^\top\biX_g$ are
singular. However, the problem solves itself if we partial out the
fixed-effect dummies and replace $\biX$ by $\tilde\biX$ and $\biy$ by
$\tilde\biy$, the matrix and vector of deviations from cluster means.
For example, the \th{gj} element of $\tilde\biy$ is $y_{gj} -
N_g^{-1}\sum_{i=1}^{N_g} y_{gi}$. Since this depends only on
observations for cluster~$g$, the jackknife CV$_{\tn3}$ estimator
\eqref{eq:jackvar} remains valid.




In \Cref{sec:report}, we discuss what quantities investigators should
report in any empirical analysis that involves cluster-robust
inference. In addition to measures of influence and leverage, these
may include measures of partial leverage (the analog of leverage for a
single coefficient) and summary statistics based on either leverage,
partial leverage, or the effective number of clusters. An example is
provided in \Cref{subsec:emplever}.




\subsection{Placebo Regressions}
\label{subsec:placebo}

An interesting way to assess the validity of alternative standard
errors is to run ``placebo regressions.'' The idea, first suggested in
\citet{BDM_2004}, is to start with a model and dataset, then generate
a completely artificial regressor at random, add it to the model, and
perform a $t$-test of significance. This is repeated a large number of
times, and the rejection frequency is observed. The artificial
regressor is often a dummy variable that is referred to as a ``placebo
law'' or ``placebo treatment.'' Using such a dummy variable is natural
because, for any level of intra-cluster correlation of the
disturbances, the intra-cluster correlation of the scores is greatest
for regressors that do not vary within clusters. However, any
artificial regressor that is not completely uncorrelated within
clusters can potentially be used.

Because a placebo regressor is artificial, we would expect valid
significance tests at level~$\alpha$ to reject the null close to
$\alpha\%$ of the time when the experiment is repeated many times.
Following the lead of \citet{BDM_2004} by using models for
log-earnings based on age, education, and other personal
characteristics, together with data taken from the Current Population
Survey, several papers \citep*{JGM_2016,MW-JAE,Brewer_2018} find that
not clustering, or clustering at below the state level, leads to
rejection rates far greater than~$\alpha$. In
\Cref{subsec:emp-placebo}, we find similar results for the datasets
used in our empirical example. Our findings, and those of the papers
cited above, all suggest that using a state\tkk-level CRVE is
important for survey data that samples individuals from multiple
states. If we fail to do so, we will find, with probability much
higher than $\alpha$, that nonsense regressors apparently belong in
the model.


Since the empirical score vectors are $\hat\bis_g =
\biX_g^\top\hat\biu_g$, a placebo\tkk-regressor experiment should lead
to over-rejection whenever both the regressor and the residuals
display intra-cluster correlation at a coarser level than the one at
which the standard errors are clustered. As in \Cref{subsec:level},
suppose there are two potential levels of clustering, fine and coarse,
with the fine clusters nested within the coarse clusters. If the
placebo regressor is clustered at the coarse level, we would expect
significance tests based on heteroskedasticity-robust standard errors
to over-reject whenever the residuals are clustered at either level.
Similarly, we would expect significance tests based on
finely-clustered standard errors to over-reject whenever the residuals
are clustered at the coarse level. \Cref{tab:placebo} in
\Cref{subsec:emp-placebo} displays both of these phenomena.

Placebo regressions can provide useful guidance as to the correct
level of clustering. However, using the rejection rates for placebo
regressions with different levels of clustering as informal tests is
really a form of pre\tkk-testing. Thus, like using the formal tests
discussed in \Cref{subsec:level}, doing this seems very likely to
yield less conservative inferences than simply relying on the second
rule of thumb.

\subsection{Two\tk-\tn Way Clustering}
\label{subsec:twoway}

Up to this point, we have assumed that there is clustering in only one
dimension. However, there could well be clustering in two or more
dimensions. With data that have both a spatial and a
temporal dimension, there may be clustering by jurisdiction and also
by time period. In finance, there is often clustering by firm and 
by year. Thus, instead of \eqref{eq:lrmodel}, we might have
\begin{equation}
\label{eq:modelgh}
\biy_{gh} = \biX_{gh} \bbeta + \biu_{gh},\quad g=1,\ldots,G,
\;\; h=1,\ldots,H,
\end{equation}
where the vectors $\biy_{gh}$ and $\biu_{gh}$ and the matrix
$\biX_{gh}$ contain, respectively, the rows of $\biy$, $\biu$, and
$\biX$ that correspond to both the \th{g} cluster in the first
clustering dimension and the \th{h} cluster in the second one. The
$GH$ clusters into which the data are divided in \eqref{eq:modelgh}
represent the intersection of the two clustering dimensions.

If there are $N_g$ observations in the \th{g} cluster for the first
dimension, $N_h$ observations in the \th{h} cluster for the second
dimension, and $N_{gh}$ observations in the \th{gh} cluster for the
intersection, the number of observations in the entire sample is
$N=\sum_{g=1}^G N_g=\sum_{h=1}^H N_h=\sum_{g=1}^G \sum_{h=1}^H
N_{gh}$, where $N_{gh}$ might equal 0 for some values of $g$ and $h$.
The scores for the clusters in the first dimension are
$\bis_g=\biX_g^\top\biu_g$, for the clusters in the second dimension
$\bis_h=\biX_h^\top\biu_h$, and for the intersections
$\bis_{gh}=\biX_{gh}^\top\biu_{gh}$. If, by analogy with
\eqref{eq:Sigma_g}, we assume that
\begin{equation}
\label{eq:varmats}
\bSigma_g = \E(\bis_g \bis_g^\top), \;
\bSigma_h = \E(\bis_h \bis_h^\top), \;
\bSigma_{gh} = \E(\bis_{gh} \bis_{gh}^\top),\;
\E(\bis_{gh} \bis_{g'h'}^\top)=\bzero \; \text{for } g\neq g', h \neq h' ,
\end{equation}
then the variance matrix of the scores is seen to be
\begin{equation}
\label{eq:scorevar}
\bSigma = \sum_{g=1}^G \bSigma_g
   + \sum_{h=1}^H \bSigma_h 
   - \sum_{g=1}^G\sum_{h=1}^H \bSigma_{gh}.
\end{equation}
The last condition in \eqref{eq:varmats} means that the scores are
assumed to be independent whenever they do not share a cluster along
either dimension. The third term in \eqref{eq:scorevar} must be
subtracted in order to avoid double counting. It is important to
distinguish between two\tkk-way clustering and clustering by the
intersection of the two dimensions. If we assumed the latter instead
of the former, then all three terms on the right-hand side of
\eqref{eq:scorevar} would be equal, and consequently $\bSigma =
\sum_{g=1}^G \sum_{h=1}^H \bSigma_{gh}$. Thus these assumptions are
radically different.


An estimator of the variance matrix of $\hat\bbeta$ is
\begin{equation}
\label{eq:betavar}
\widehat\var(\hat\bbeta) =
(\biX^\top\biX)^{-1} \hat\bSigma (\biX^\top\biX)^{-1}\tn, \quad
\hat\bSigma = \sum_{g=1}^G \hat\bis_g\hat\bis_g^\top +
\sum_{h=1}^H \hat\bis_h\hat\bis_h^\top -
\sum_{g=1}^G\sum_{h=1}^H \hat\bis_{gh}\hat\bis_{gh}^\top.
\end{equation}
Here $\hat\bSigma$ is an estimator of \eqref{eq:scorevar}, with the
empirical scores defined in the usual way; for example, $\hat\bis_g =
\biX_g^\top\hat\biu_g$. In practice, each of the matrices on the
right-hand side of the second equation in \eqref{eq:betavar} is
usually multiplied by a scalar factor, like the one in \eqref{eq:CV1},
designed to correct for degrees of freedom. Because the third term is
subtracted, the matrix $\hat\bSigma$ may not always be positive
definite. This problem can be avoided by omitting the third term,
which is asymptotically valid under some assumptions
\citep*{Davezies_2021,MNW_2021}. Another possibility is to use an
eigenvalue decomposition \citep*{CGM_2011}, although this merely
forces the variance matrix to be positive semidefinite.

The idea of two\tkk-way clustering can, of course, be generalized to
three\tkk-way clustering, four-way clustering, and so on. However, the
algebra rapidly becomes daunting. If there were three clustering
dimensions, for example, the analog of \eqref{eq:scorevar} would have
seven terms.

Two\tkk-way clustering seems to have been suggested first in
\citet{MH_2006} and rediscovered independently by \citet*{CGM_2011}
and \citet{Thompson_2011}. Although two\tkk-way clustering has been
widely used in empirical work, the asymptotic theory to justify it is
much more challenging than the theory for the one\tkk-way case, and
this theory is still under active development
\citep*{Chiang_2020,Chiang_2021JBES,Davezies_2021,MNW_2021,Menzel_2021}.
In view of this, and because of the technical difficulties involved,
we will focus mainly on one\tkk-way clustering in the remainder of the
paper.

\section{Asymptotic Inference}
\label{sec:asytheory}


For the regression model \eqref{eq:lrmodel}, inference is commonly based 
on the $t$-statistic,
\begin{equation}
t_a = \frac{\bia^\top(\hat\bbeta -
\bbeta_0)}{(\bia^\top\hat\biV\bia)^{1/2}}\tk,
\label{eq:tstat}
\end{equation}
where the hypothesis to be tested is
$\bia^\top\bbeta=\bia^\top\bbeta_0$, with $\bia$ a known $k$-vector. 
Here $\hat\biV$ denotes one of CV$_{\tn1}$, CV$_{\tn2}$, or
CV$_{\tn3}$, given in \eqref{eq:CV1}, \eqref{eq:CV2}, and either
\eqref{eq:CV3} or \eqref{eq:jackvar}, respectively. In many cases,
just one element of $\bia$, say the \th{j}, equals~1, and the
remaining elements equal~0, so that \eqref{eq:tstat} is simply
$\hat\beta_j - \beta_{j0}$ divided by its standard error. When there
are $r>1$ linear restrictions, which can be written as $\biR\bbeta =
\bir$ with $\biR$ an $r\times k$ matrix, inference can be based on the
Wald statistic,
\begin{equation}
W = (\biR\hat\bbeta - \bir)^\top
(\biR\tkk\hat\biV\tn\biR^\top)^{-1} (\biR\hat\bbeta - \bir).
\label{Waldstat}
\end{equation}
Of course, when $r=1$, the $t$-statistic \eqref{eq:tstat} is just the 
signed square root of a particular Wald statistic with
$\biR=\bia^\top$ and $\bir = \bia^\top\bbeta_0$.

By letting the sample size become arbitrarily large, one can
frequently obtain a tractable asymptotic distribution for any test
statistic of interest, including \eqref{eq:tstat} and
\eqref{Waldstat}. Ideally, this would provide a good approximation to
the actual distribution. With clustered data, there is more than one
natural way to let the sample size become large, because we can make
various assumptions about what happens to $G$ and the $N_g$ as we let
$N$ tend to infinity. Which assumptions it is appropriate to use, and
how well the resulting approximations work, will depend on the
characteristics of the sample and the (unknown) DGP.

In order for inferences based on the statistics \eqref{eq:tstat} and
\eqref{Waldstat} to be asymptotically valid, two key asymptotic
results must hold. First, a central limit theorem (CLT) must apply to
the sum of the score vectors $\bis_g$ in \eqref{eq:betahat}. In the
limit, after appropriate normalization, the vector $\sum_{g=1}^G
\bis_g$ needs to follow a multivariate normal distribution with
variance matrix $\sum_{g=1}^G \bSigma_g$. Second, again after
appropriate normalization, a law of large numbers (LLN) must apply to
the matrices $\sum_{g=1}^G \hat\bis_g\hat\bis_g^\top$, $\sum_{g=1}^G
\grave\bis_g\grave\bis_g^\top$, or $\sum_{g=1}^G
\acute\bis_g\acute\bis_g^\top$ in the middle of the variance matrix
estimators \eqref{eq:CV1}, \eqref{eq:CV2}, or \eqref{eq:CV3}, so that
they converge to $\sum_{g=1}^G \bSigma_g$. We refer to ``appropriate
normalization'' here rather than specifying the normalization factors
explicitly because, with clustered data, the issue of normalization is
a very tricky one; see \Cref{subsec:large}. For asymptotic inference
to be reliable, we need both the CLT and the LLN to provide good
approximations.


There are currently two quite different types of assumptions on which
the asymptotic theory of cluster-robust inference can be based. The
most common approach, and we believe usually the most appropriate one,
is to let the number of clusters tend to infinity. We refer to this as
the ``large number of clusters'' approach and discuss it in
\Cref{subsec:large}. An alternative approach is to hold the number of
clusters fixed and let the number of observations within each cluster
tend to infinity. We refer to this as the fixed-$G$ or ``small number
of large clusters'' approach and discuss it in \Cref{subsec:small}.
Some of the material in \Cref{subsec:large,subsec:small} is quite
technical, but it helps to explain when and why asymptotic inference
can fail.

Inference based on asymptotic theory often performs well, but it can
perform poorly in some commonly-encountered situations that are
discussed in \Cref{subsec:failure}. We therefore do not recommend
relying only on CV$_{\tn1}$ and asymptotic theory. Because the
bootstrap methods to be discussed in \Cref{sec:bootstrap} can work
much better than asymptotic methods when the latter do not work well,
we recommend that they be used almost all the time, at least to verify
that both approaches yield similar results. In particular, we
recommend using one or more variants of the wild cluster restricted,
or WCR, bootstrap (\Cref{subsec:wild}) as a matter of routine.





\subsection{Asymptotic Theory: Large Number of Clusters}
\label{subsec:large}


The simplest assumption about how the sample size goes to infinity is
that every cluster has a fixed number of observations, say $M$\tn.
Then $N=MG$, and both $N$ and $G$ go to infinity at the same rate.
Thus the appropriate normalizing factor for the parameter estimator is
either $\sqrt{G}$ or $\sqrt{N}$. In this case, it is not difficult to
show that $\sqrt{G}(\hat\bbeta - \bbeta_0)$ is asymptotically
multivariate normal with variance matrix equal to the probability
limit of $G$ times the right-hand side of \eqref{eq:trueV}. Moreover,
the latter can be estimated consistently by $G$ times the CV$_{\tn1}$,
CV$_{\tn2}$, or CV$_{\tn3}$ matrices. The first proof for this case of
which we are aware is in \citet[Chapter~6]{White_1984}; see also
\citet{Hansen_2007}.


In actual samples, clusters often vary greatly in size, so it is
usually untenable to assume that every cluster has the same number of
observations. The assumption that $G$ is proportional to $N$ may be
relaxed by allowing $G$ to be only approximately proportional to
$N$\tn, so that $G/N$ is roughly constant as $N\to\infty$. This
implies that all the clusters must be small. In this case, the quality
of the asymptotic approximations is not likely to be harmed much by
moderate variation in cluster sizes. If a sample has, say, 500
clusters that vary in size from 10 to 50 observations, we would expect
asymptotic inference to perform well unless there is some other reason
(unrelated to cluster sizes) for it to fail.

\citet*{DMN_2019} and \citet{HansenLee_2019} take a more flexible 
approach, with primitive conditions that restrict the variation in the
$N_g$ relative to the sample size. These conditions allow some
clusters to be ``small'' and others to be ``large'' in the sense that
some but not all $N_g \to \infty$ as $N \to \infty$. Although a key
assumption is that $G\to\infty$ (i.e., is ``large''), the appropriate
normalization factor for $\hat\bbeta - \bbeta_0$ is usually not
$\sqrt{G}$. Instead, this factor depends in a complicated way on the
regressors, the relative cluster sizes, the intra-cluster correlation
structure, and interactions among these; some examples of different
normalizing factors are given in the papers cited above. For this
reason, the key result that the $t$-statistic defined in
\eqref{eq:tstat} is asymptotically distributed as standard normal is
derived assuming that the rate at which $\hat\bbeta - \bbeta_0$ tends
to zero is unknown. Of course, this result also justifies using the
$t(G-1)$ distribution, which is more conservative, is widely used, and
is derived from the theory discussed in \Cref{subsec:small}.

The application of a CLT to $\sum_{g=1}^G \bis_g$, appropriately
normalized, requires a restriction on the amount of heterogeneity that
is allowed. Otherwise, just a few clusters might dominate the entire
sample in the limit, thus violating the Lindeberg or Lyapunov
conditions for the CLT. The necessary restrictions on the
heterogeneity of clusters may be expressed in terms of two key
parameters. The first of these parameters is the number of moments
that is assumed to exist for the distributions of $\bis_{gi}$
(uniformly in $g$ and~$i$). We denote this parameter by $\gamma > 2$.
When more moments exist, the distributions of $\bis_{gi}$ are closer
to the normal distribution, and hence the sample will feature fewer
outliers or other highly leveraged observations or clusters.

In the clustered regression model, the variance of the scores, which
is often referred to as the Fisher information matrix, is given by
$\mathcal{J}_N = \sum_{g=1}^G \var (\bis_g )$. When appropriately
normalized, $\mathcal{J}_N$ converges to a nonzero and finite matrix
$\mathcal{J}$\tn. The rate of convergence $\eta_N$ is defined
implicitly by $\eta_N^{-1} \mathcal{J}_N \to \mathcal{J}$\tn. This
rate is the second key parameter. One interpretation of $\eta_N$ can
be found in \eqref{eq:trueV}, from which the stochastic order of
magnitude of $\hat\bbeta - \bbeta_0$ is seen to be
$O_P(\eta_N^{1/2}\!/N)$. In general, $\eta_N \geq N$, with the
equality holding whenever there is no intra-cluster correlation. The
larger the value of $\eta_N$, the more slowly does $\hat\bbeta$
converge to $\bbeta_0$.

The conditions required on the heterogeneity of clusters to apply a 
CLT can be stated in terms of the parameters $\gamma$ and $\eta_N$.
Specifically, when expressed in our notation, Assumption~3 of
\citet{DMN_2019} states the following condition:
\begin{equation}
\label{RateCondition}
\Bigl(\frac{\eta_N^{1/2}}{N}\Bigr)^{\!\frac{-2\gamma}{2\gamma-2}}
\,\frac{\sup_g N_g}{N} \longto 0.
\end{equation}
Because $\eta_N = o ( N^2)$ for consistency of $\hat\bbeta$, the 
condition in \eqref{RateCondition} makes it clear that we cannot allow
a single cluster to dominate the sample, in the sense that its size is
proportional to~$N$\tn. More generally, \eqref{RateCondition} shows
that there is a tradeoff between information accumulation and
variation in cluster sizes, as measured by the largest cluster size.
To interpret this tradeoff, we will consider three different
implications of \eqref{RateCondition}.


First, when $\gamma$ increases, so that more moments exist, condition
\eqref{RateCondition} becomes less strong. In particular, when the
scores are nearly normally distributed, in the sense that all their
moments exist, then $\gamma=\infty$, which implies that
$-2\gamma/(2\gamma-2) = -1$. In this case, \eqref{RateCondition}
reduces to $\eta_N^{-1/2} \sup_g N_g \to 0$, so that the size of the
largest cluster must increase more slowly than the square root of the
rate at which the Fisher information matrix converges. When
$\gamma<\infty$, so that there are fewer moments, then the rate at
which $\sup_g N_g$ is allowed to increase becomes smaller.

Second, suppose that the scores are uncorrelated, or more generally
that a CLT applies to $N_g^{-1/2}\bis_g$, as assumed in
\citet*{BCH_2011} and \citet{IM_2010,IM_2016}; see
\Cref{subsec:small}. In this case, $\var (\bis_g)= O(N_g)$, so that
$\eta_N = N$. Condition \eqref{RateCondition} is then
$N^{-(\gamma-2)/(2\gamma-2)} \sup_g N_g \to 0$. If the scores are
nearly normal, then it reduces further to $N^{-1/2} \sup_g N_g \to 0$,
so that the size of the largest cluster must increase no faster than
the square root of the sample size. Once again, the fewer moments
there are, the more slowly $\sup_g N_g$ is allowed to increase.

Third, suppose that the scores are generated by the factor model in 
\eqref{eq:factor}, or by the simpler random-effects model
\eqref{eq:REmodel}. Then $\var (\bis_g)=O(N_g^2)$. If, in addition,
$\inf_g N_g$ and $\sup_g N_g$ are of the same order of magnitude, then
$\eta_N = N \sup_g N_g$ and the condition \eqref{RateCondition}
collapses to $N^{-1} \sup_g N_g \to~0$, regardless of the number of
finite moments.


One possibly surprising implication of the above discussion is that,
when there is more intra-cluster correlation, so that $\eta_N$ is
relatively large, then greater heterogeneity of cluster sizes is
allowed. That is, a higher degree of intra-cluster correlation implies
a faster rate of convergence, $\eta_N$, of the Fisher information 
matrix, which in turn allows a larger $\sup_g N_g$ in
\eqref{RateCondition}. The intuition is that greater intra-cluster
correlation reduces the effective cluster size, as measured by the
amount of independent information a cluster contains. In the extreme
case in which all observations in the \th{g} cluster are perfectly
correlated, the size of the cluster is effectively~1 and not $N_g$.
Note, however, that large clusters are implicitly weighted more
heavily than small clusters even in this extreme case.

Although it is impossible to verify condition \eqref{RateCondition} in
any finite sample, investigators can always observe the $N_g$. The 
discussion of \eqref{RateCondition} above suggests that asymptotic 
inference tends to be unreliable when the $N_g$ are highly variable,
especially when a very few clusters are unusually large. This is
exacerbated when the distribution of the data is heavy-tailed (has
fewer moments), but is mitigated when the clusters have an 
approximate factor structure. Circumstances in which asymptotic
inference can be unreliable are discussed in more detail in
\Cref{subsec:failure}.

\subsection{Asymptotic Theory: Small Number of Large Clusters}
\label{subsec:small}

A few authors have assumed that $G$ remains fixed (i.e., is ``small'')
as $N\to\infty$, while the cluster sizes diverge (i.e., are
``large''). Notably, \citet{BCH_2011} proves that, for CV$_{\tn1}$,
the $t$-statistic \eqref{eq:tstat} follows the $t(G-1)$ distribution
asymptotically. An analogous result for the Wald statistic
\eqref{Waldstat} is discussed in \Cref{subsubsec:several}. These
results provide useful approximations, but they are proven under some
very strong assumptions. In particular, all the clusters are assumed
to be the same size $M$\tn. In addition, the pattern of dependence
within each cluster is assumed to be such that a CLT applies to the
normalized score vectors $M^{-1/2}\bis_g$ for all $g=1,\ldots,G$, as
$M\to\infty$.


This second assumption is crucial, as it limits the amount of
dependence within each cluster and requires it to diminish quite
rapidly as $M\to\infty$. Although \citet{BCH_2011} discusses a 
particular model for which this requirement holds, it rules out simple
DGPs such as the factor model \eqref{eq:factor}, with or without
cluster fixed effects. It even rules out the random-effects model 
\eqref{eq:REmodel}, which is the most common model of intra-cluster 
correlation. For all these models, no CLT can possibly apply to the 
vector~$M^{-1/2} \bis_g$. Consider, for example, the factor model 
\eqref{eq:factor}, and suppose that $\bix_g=1$. In this case,
\begin{equation}
\label{clt:variance}
\var ( M^{-1/2} \bis_g ) = \frac1M\!\sum_{i,j=1}^{M}\!\cov (u_{gi},u_{gj})
= \frac1M\sum_{i=1}^{M} \lambda_{gi}^2
+ \frac2M\sum_{i=1}^{M} \sum_{j=i+1}^{M}\!\lambda_{gi}\lambda_{gj}.
\end{equation}
Because of the double summation, the second term on the right-hand side 
of \eqref{clt:variance} clearly does not converge as $M \to \infty$
unless additional, and very strong, assumptions are made.





Another, quite different, approach to inference when $G$ is fixed is
developed in \citet{IM_2010}. The parameter of interest is a scalar,
say $\beta$, which can be thought of as one element of $\bbeta$. The
key idea is to estimate $\beta$ separately for each of the $G$
clusters. This yields estimators $\hat\beta_g$ for $g=1,\ldots,G$.
Inference is then based on the average, say $\bar\beta$, and standard
error, say $s_{\tn\hat\beta}$, of the $\hat\beta_g$. \citet{IM_2010}
shows that the test statistic $\sqrt{G}(\bar\beta-\beta_0)
/s_{\tn\hat\beta}$ is approximately distributed as $t(G-1)$ when all 
clusters are large and a CLT applies to $N_g^{-1/2} \bis_g$ for 
each~$g$. However, as we saw above, this assumption cannot hold even
for the simple random-effects or factor models.

A practical problem with this procedure is that $\beta$ may not be
estimable for at least some clusters. For models of treatment effects
at the cluster level, this will actually be the case for every
cluster. For difference\tkk-in-differences (DiD) models with
clustering at the jurisdiction level, it will be the case for every
jurisdiction that is never treated. \citet{IM_2016} suggests a way to
surmount this problem by combining clusters into larger ones that
allow $\beta$ to be estimated for each of them. Even when $\beta$
itself can be estimated for each cluster, however, the full model may
not be estimable. This can happen, for example, when there are fixed
effects for categorical variables, and not all categories occur in
each cluster. In such cases, the interpretation of $\beta$ may differ 
across clusters.


Although the estimators and test statistics proposed in
\citet{IM_2010,IM_2016} differ from the more conventional ones studied
in \citet{BCH_2011}, both approaches lead to $t$-statistics that
follow the $t(G-1)$ distribution asymptotically. This distribution has
in fact been used in \texttt{Stata} as the default for
CV$_{\tn1}$-based inference for many years. For small values of $G$,
using it can lead to noticeably more accurate, and more conservative,
inferences than using the $t(N-k)$ or normal distributions. However,
as we discuss in the next subsection and in \Cref{sec:bootstrap},
inferences based on the $t(G-1)$ distribution are often not nearly
conservative enough, especially when $G$ is small.



\subsection{When Asymptotic Inference Can Fail}
\label{subsec:failure}


Whenever we rely on asymptotic theory, we need to be careful. What is
true for infinitely large samples may or may not provide a good
approximation for any actual sample. Unless the very strong
assumptions discussed in \Cref{subsec:small} are satisfied, we cannot
expect to obtain reliable inferences when $G$ is small.

Unfortunately, there is no magic number for $G$ above which asymptotic
inference can be relied upon. It is sometimes claimed that asymptotic
inference based on CV$_{\tn1}$ is reliable when $G \geq 50$, or even
(partly in jest) when $G \geq 42$ \citep{MHE_2008}, but this is not
true. In very favorable cases, inference based on CV$_{\tn1}$ and the
$t(G-1)$ distribution can be fairly reliable when $G=20$, but in
unfavorable ones it can be unreliable even when $G=200$ or more; see
\Cref{subsubsec:het,subsubsec:few}. Moreover, there is evidence
\citep{BM_2002,MNW-bootknife} that inference based on CV$_{\tn3}$ 
tends to be more reliable, sometimes much more reliable, than
inference based on other CRVEs.





\subsubsection{Cluster Heterogeneity}
\label{subsubsec:het}


What determines whether a case is favorable or unfavorable for a given
$G$ is mostly the heterogeneity of the cluster score vectors.
Unfortunately, the latter cannot be observed directly. We can observe
the empirical score vectors, but they can sometimes differ greatly
from the true ones (\Cref{subsubsec:few}). We can also observe cluster
sizes, which the discussion in \Cref{subsec:large} focused on. These
are often particularly important, but any form of heterogeneity can
have serious consequences. This includes both heteroskedasticity of
the disturbances at the cluster level and systematic variation across
clusters in the distribution of the regressors. In general, the number
of clusters $G$ and the extent to which the distribution of the scores
varies across clusters will determine the quality of the asymptotic
approximation.




In principle, a poor asymptotic approximation could lead $t$-tests
based on the $t(G-1)$ distribution either to under-reject or
over-reject. We have never observed $t$-tests based on CV$_{\tn1}$ or
CV$_{\tn2}$ to under-reject in any simulation experiments, but we have
observed ones based on CV$_{\tn3}$ to do so. Unless $G$ is fairly
large, it is not difficult to find cases in which \mbox{$t$-tests}
based on any of these CRVEs reject much more than their nominal level.
Wald tests of several restrictions typically perform even worse; see
\Cref{subsubsec:several}.

As we discussed in \Cref{subsec:large}, the condition
\eqref{RateCondition} imposes a restriction on the size of the largest
cluster relative to the sample size. Thus, the quality of the 
asymptotic approximation will surely diminish as the size of the
largest cluster increases relative to the average cluster size, and 
over-rejection will consequently increase. This conjecture is
supported by simulation evidence in \citet{MW-JAE} and
\citet{DMN_2019}, as well as by analytic results based on Edgeworth
expansions in the latter paper.

There are at least two situations in which cluster-robust $t$-tests
and Wald tests are at risk of over-rejecting to an extreme extent,
namely, when one or a few clusters are unusually large, or when only a
few clusters are treated. In both of these cases, one cluster, or just
a few of them, have high leverage, in the sense that omitting one of
these clusters has the potential to change the OLS estimates
substantially; see \Cref{subsec:leverage}. Since both of these
situations can occur even when $G$ is not small, all users of
cluster-robust inference need to be on guard for them.


The first case in which conventional inference fails is when one or a
very few clusters are much larger than the others. This implies that
the distributions of the score vectors for those clusters are much
more spread out than the ones for the rest of the clusters. An extreme
example is studied in \citet[Figure~3]{DMN_2019}. When half the sample
is in one large cluster, rejection rates for $t$-tests based on
CV$_{\tn1}$ actually increase as $G$ increases, approaching 50\% at
the 5\% level for $G=201$. Unfortunately, this extreme case is
empirically relevant. Because roughly half of all incorporations in
the United States are in Delaware, empirical studies of state laws and
corporate governance encounter precisely this situation whenever they
cluster at the state level \citep{HS_2020}. \citet{MNW-bootknife}
studies a case with cluster sizes proportional to incorporations,
where almost 53\% of the observations are in the largest cluster.
Although $t$-tests based on all three CRVEs over-reject, those based
on CV$_{\tn3}$ do so much less severely than ones based on CV$_{\tn1}$
and CV$_{\tn2}$.


Not all forms of heterogeneity are harmful. In particular, having some
extremely small clusters in a sample generally does not cause any
problems, so long as there is not too much heterogeneity in the
remainder of the sample. For example, suppose that a sample consists
of, say, 25 large clusters, each with roughly 200 observations, and 15
tiny clusters, each with just one or a handful of observations. Except
in very unusual cases, the coefficient estimates and their
$t$-statistics would hardly change if we were to drop the tiny
clusters, so this sample is better thought of as having 25 equal-sized 
clusters. The asymptotic approximations would perform just about the 
same whether or not the tiny clusters were included. Of course, if we
changed this example so that there were 5 large clusters and 15 tiny
ones, then asymptotic inference would surely be very problematic,
because there would effectively be just 5 clusters.




\subsubsection{Treatment and Few Treated Clusters}
\label{subsubsec:few}

The second case in which conventional inference fails is when the
regressor of interest is a treatment dummy, and treatment occurs only
for observations in a small number of clusters. In such cases, the
empirical score vectors for the treated clusters, even when they have
been modified, can provide very poor estimates of the actual score
vectors.


Suppose that $d_{gi}$ is the value of the treatment dummy for
observation $i$ in cluster $g$, and let $s^d_g$ denote the element of
$\bis_g$ corresponding to the dummy. Consider first the extreme case
in which only some or all of the observations in the first cluster are
treated. Then $s^d_g = \sum_{i=1}^{N_g} d_{gi} u_{gi}$ is equal to
$\sum_{i=1}^{N_1} d_{1i} u_{1i}$ for $g=1$ and to~0 for all $g \neq
1$. Thus the scores corresponding to the treatment dummy equal zero
for the control clusters. Moreover, because the treatment regressor
must be orthogonal to the residuals, the empirical score $\hat s^d_1 =
0$. Since the actual score $s^d_1 \neq 0$, this implies that
\eqref{eq:CV1} provides a dreadful estimate of \eqref{eq:trueV}, at
least for the elements corresponding to the coefficient on the
treatment dummy. In consequence, the CV$_{\tn1}$ standard error of
this coefficient can easily be too small by a factor of five or more.
When more than one cluster is treated, the problem is not as severe,
because the $\hat s^d_g$ now sum to zero over the observations in all
the treated clusters. This causes them to be too small, but not to the
same extent as when just one cluster is treated; see
\citet{MW-JAE,MW-EJ}.

How well the empirical scores mimic the actual scores depends on the
sizes of the treated and control clusters, the values of other
regressors, and the number of treated observations within the treated
clusters. Thus all these things affect the accuracy of cluster-robust
standard errors and the extent to which $t$-statistics based on them
over-reject. As the number of treated clusters, say $G_1$, increases,
the problem often goes away fairly rapidly. But increasing $G$ when
$G_1$ is small and fixed does not help and may cause over-rejection to
increase. For models where all observations in each cluster are either
treated or not, having very few control clusters is just as bad as
having very few treated clusters. The situation is more complicated
for DiD models, however; see \citet{MW-TPM}.

It seems plausible that the modified empirical score vectors used by
CV$_{\tn2}$ and CV$_{\tn3}$ may mimic the actual score vectors more
accurately than the unmodified ones used by CV$_{\tn1}$. In fact,
there is some evidence \citep{MNW-bootknife} that $t$-tests based on
CV$_{\tn2}$ over-reject less severely than ones based on CV$_{\tn1}$
when there are few treated clusters, and that $t$-tests based on
CV$_{\tn3}$ over-reject considerably less severely. However, even
though CV$_{\tn3}$ can perform much better than CV$_{\tn1}$, it 
still tends to over-reject when $G_1$ is very small.




\subsubsection{Testing Several Restrictions}
\label{subsubsec:several}


Most of the literature on cluster-robust inference has focused on
$t$-tests, but the cluster-robust Wald tests defined in
\eqref{Waldstat} also generally over-reject in finite samples. In
fact, they tend to do so more severely as $r$, the number of
restrictions, increases; see \citet{PT_2018} and \citet{JGM-fast}. As
is well known, this phenomenon occurs for Wald tests of all kinds. The
problem may well be unusually severe in this case, however, because
all CRVEs have rank at most $G$ and, in many cases, only $G-1$. On the
other hand, although the true variance matrix $\var(\hat\bbeta)$ is
unknown in finite samples, it will normally have full rank~$k$. As $r$
increases, the inverse of $\biR^\top\hat\biV\biR$ thus seems likely to
provide an increasingly poor approximation to the inverse of
$\biR^\top\!\var(\hat\bbeta)\biR$\tkk.



\citet{BCH_2011} proves a result that helps to mitigate this problem.
Under the very strong assumptions discussed in \Cref{subsec:small},
where $G$ is fixed, every cluster is the same size $M$, and the amount
of within-cluster dependence is limited, the paper shows that
$r(G-1)/(G-r)$ times the appropriate quantile of the $F(r,G-r)$
distribution provides an asymptotic critical value for the Wald
statistic \eqref{Waldstat} based on CV$_{\tn1}$. Equivalently,
$(G-r)/(r(G-1))$ times $W$ is shown to follow the $F(r,G-r)$
distribution asymptotically in this special case. However, the WCR
bootstrap (\Cref{subsec:wild}) can provide a much better
approximation.



\subsection{Cluster-Robust Inference in Nonlinear Models}
\label{subsec:nonlinear}

Although cluster-robust inference is most commonly used with the
linear regression model \eqref{eq:lrmodel}, it can actually be
employed for a wide variety of models estimated by maximum likelihood
or the generalized method of moments (GMM); see
\citet{HansenLee_2019}.

Consider a model characterized by the log-likelihood function
\begin{equation}
\ell(\btheta) = \sum_{g=1}^G \sum_{i=1}^{N_g} \ell_{gi}(\btheta),
\label{logl}
\end{equation}
where $\btheta$ is the $k\times 1$ parameter vector to be estimated,
and $\ell_{gi}(\btheta)$ denotes the contribution to the log-likelihood
made by the \th{i} observation within the \th{g} cluster. Let
$\hat\btheta$ denote the vector that maximizes \eqref{logl},
$\bis_{gi}(\btheta)$ the $k\times 1$ vector of the first derivatives
of $\ell_{gi}(\btheta)$ (that is, the score vector), and
$\biH_{gi}(\btheta)$ the $k\times k$ Hessian matrix of the second
derivatives. Further, let $\hat\bis_g = \sum_{i=1}^{N_g}
\bis_{gi}(\hat\btheta)$ and $\hat\biH = \sum_{g=1}^G \sum_{i=1}^{N_g}
\biH_{gi}(\hat\btheta)$. Then \citet[Theorem~10]{HansenLee_2019} shows
(using somewhat different notation) that the cluster-robust variance
estimator for the maximum likelihood estimator $\hat\btheta$ is
\begin{equation}
\widehat\var(\hat\btheta - \btheta_0) =
\hat\biH^{-1} \bigg(\sum_{g=1}^G \hat\bis_g \hat\bis_g^\top\!\bigg)
\tk\hat\biH^{-1}.
\label{mlvar}
\end{equation}
The resemblance between \eqref{mlvar} and the CV$_{\tn1}$ variance
matrix in \eqref{eq:CV1} is striking. Indeed, since the Hessian is
proportional to $\biX^\top\biX$ for the linear regression model,
CV$_{\tn1}$ without the leading scalar factor is really just a special
case of \eqref{mlvar}.

The variance matrix estimator \eqref{mlvar} can be used for a wide
variety of models estimated by maximum likelihood. In fact,
\texttt{Stata} has been using it for various models, including probit
and logit, for some years. \citet[Theorem~12]{HansenLee_2019} provides
a similar result for GMM estimation, which is also very widely
applicable. More recently, the fixed-$G$ approach discussed in
\Cref{subsec:small} has been applied to GMM estimation by
\citet{Hwang_2021}. It leads to a novel inferential procedure that
involves modifying the usual asymptotic $t$ and $F$ statistics, but it
requires that cluster sizes be approximately equal.

Unfortunately, very little currently seems to be known about the
finite\tkk-sample properties of tests based on \eqref{mlvar} or its
GMM analog. They are probably worse than those of tests based on
\eqref{Waldstat}. It seems quite plausible that bootstrapping, to
which we turn in the next section, would help, and \citet{Hwang_2021}
discusses one bootstrap method for GMM estimation. However,
bootstrapping nonlinear models tends to be computationally expensive,
and the properties of applicable bootstrap procedures are largely
unknown at the present time.



\section{Bootstrap Inference}
\label{sec:bootstrap}

Instead of basing inference on an asymptotic approximation to the
distribution of a statistic of interest, it is often more reliable to
base it on a bootstrap approximation. In \Cref{subsec:bootstrap}, we
briefly review some key concepts of bootstrap testing and bootstrap
confidence intervals. Then, in \Cref{subsec:pairs,subsec:wild}, we
discuss bootstrap methods for regression models with clustered data.
These methods, in particular the wild cluster restricted (WCR)
bootstrap to be discussed in \Cref{subsec:wild}, can be surprisingly 
inexpensive to compute and often lead to much more reliable inferences
than asymptotic procedures. We therefore recommend that at least one
variant of the WCR bootstrap be used almost all the time.


\subsection{General Principles of the Bootstrap}
\label{subsec:bootstrap}

Suppose we are interested in a test statistic $\tau$, which might be a
$t$-statistic or a Wald statistic. Instead of using $P$~values or
critical values taken from an asymptotic distribution, we can use ones
from the empirical distribution function (EDF) of a large number of
bootstrap test statistics. This EDF often provides a good
approximation to the unknown distribution of~$\tau$. In order to
obtain the EDF, we need to generate $B$ bootstrap samples and use each
of them to compute a bootstrap test statistic, say $\tau^*_b$, for
$b=1,\ldots,B$.

Precisely how the bootstrap samples are generated is critical, and we
will discuss some methods for doing so in the next two subsections.
The choice of $B$ also matters. Ideally, it should be reasonably large
\citep{DM_2000} and satisfy the condition that $\alpha(B+1)$ is an
integer for any $\alpha$ (the level of the test) that may be of
interest \citep{RM_2007}. In general, the computational cost of
generating the bootstrap test statistics is proportional to $B$ times
$N$\tn, so that bootstrapping can be expensive. However, as we discuss
in \Cref{subsec:wild,subsec:pairs}, surprisingly inexpensive methods
are available for linear regression models with clustered
disturbances. Unless computational cost is an issue, $B=9,\tn999$ and
even $B=99,\tn999$ are generally good choices.


The EDF of the $\tau^*_b$ often provides a better approximation to
$F(\tau)$, the distribution of~$\tau$, than does its asymptotic
distribution. This can sometimes be shown formally, but generally only
under strong assumptions and at the cost of a great deal of algebra
\citep[Section~5]{DMN_2019}. For the model \eqref{eq:lrmodel},
however, the intuition is quite simple. In many cases, the poor
finite\tkk-sample properties of test statistics based on CV$_{\tn1}$
arise because $\sum_{g=1}^G \hat\bis_g\hat\bis_g^\top$, provides a
poor approximation to $\sum_{g=1}^G\bSigma_g$. Often the bootstrap
analog of the former provides a similarly poor approximation to the
bootstrap analog of the latter. If so, then it is plausible that the
empirical distribution of the $\tau^*_b$ will differ from the
asymptotic distribution of the $\tau^*_b$ in roughly the same way as
the distribution of~$\tau$ differs from its asymptotic distribution.
In that case, the EDF of the bootstrap test statistics should provide
a reasonably good approximation to $F(\tau)$.

The EDF of the $\tau^*_b$ may be obtained by sorting the $\tau^*_b$
from smallest to largest. Number $(1-\alpha)(B+1)$ then provides an
estimate of the $1-\alpha$ quantile, which may be used as the critical
value for an upper-tail test at level~$\alpha$. Identical inferences
will be obtained by calculating the upper-tail bootstrap $P$~value,
\begin{equation}
\hat{P}^*(\tau) = \frac{1}{B} \sum_{b=1}^B \IF(\tau^*_b > \tau),
\label{bootp}
\end{equation}
and rejecting the null hypothesis whenever $\hat{P}^*(\tau) < \alpha$.
Here $\tau$ could be either the Wald statistic \eqref{Waldstat} or the
absolute value of the $t$-statistic \eqref{eq:tstat}.

Setting $\tau = |t_a|$ in \eqref{bootp} imposes symmetry on the
bootstrap distribution of $t_a$. In many cases, it makes sense to do
this, because cluster-robust $t$-statistics for linear regression
models with exogenous regressors are often symmetrically distributed
around the origin, at least to a good approximation. When $\tau =
|t_a|$, the quantity $\hat{P}^* (\tau)$ defined in \eqref{bootp} is a
symmetric bootstrap $P$~value for a two\tkk-sided test of
$\bia^\top(\bbeta-\bbeta_0)=0$; see~\eqref{eq:tstat}.

In dynamic models, nonlinear models, and models estimated by
instrumental variables, however, it is common for the coefficients of
interest to be biased. This causes the associated $t$-statistics to
have non-zero means in finite samples. In such cases, it makes sense
to use the equal-tail bootstrap $P$~value,
\begin{equation}
\hat{P}^*_{\rm et}(\tau) = \frac{2}{B}\min\Big(\sum_{b=1}^B\IF(\tau^*_b >
\tau),\; \sum_{b=1}^B\IF(\tau^*_b \le \tau)\tn\Big).
\label{equaltail}
\end{equation}
Here we compute upper-tail and lower-tail $P$~values, take the minimum
of them, and then multiply by 2 to ensure that the nominal level of the
test is correct.

There are many ways to construct a bootstrap confidence interval for a
regression coefficient $\beta$ of which we have an estimate
$\hat\beta$. A method that is conceptually (but not always
computationally) simple is to invert a bootstrap test. This means
finding two values of the coefficient, say $\beta_l$ and $\beta_u$,
with $\beta_u > \beta_l$ and normally on opposite sides of
$\hat\beta$, such that
\begin{equation}
\hat{P}^*_{\rm et}\big(t(\beta=\beta_l)\big) = \alpha
\quad\mbox{and}\quad
\hat{P}^*_{\rm et}\big(t(\beta=\beta_u)\big) = \alpha.
\label{eq:confint}
\end{equation}
Here $t(\beta=\beta_c)$, for $c=l$ and $c=u$, is a cluster-robust
$t$-statistic for the hypothesis that $\beta=\beta_c$. The desired
$1-\alpha$ confidence interval is then $[\beta_l,\, \beta_u]$. When
the bootstrap DGP imposes the null hypothesis, the distribution of the
bootstrap samples depends on the value of $\beta_c$. Solving the two
equations in \eqref{eq:confint} therefore requires iteration; see
\citet{Hansen_1999} and \citet{JGM_2015,JGM-fast}. In general, these
methods tend to be expensive, but this is not the case for the WCR
bootstrap to be discussed in \Cref{subsec:wild}.


For bootstrap confidence intervals, it is common to use bootstrap DGPs 
that do not impose the null hypothesis, because no iteration is then 
required. The simplest method is just to calculate the standard 
deviation of the $\hat\beta^*_b$ and use this number, say 
$s^*(\hat\beta)$, as an estimator of the standard error of $\hat\beta$. 
The confidence interval is then
\begin{equation}
\label{eq:sstarint}
\big[\hat\beta - c_{1-\alpha/2}\, s^*(\hat\beta), \;
\hat\beta + c_{1-\alpha/2}\, s^*(\hat\beta)\big],
\end{equation}
where $c_{1-\alpha/2}$ is the $1 - \alpha/2$ quantile of (in this
case) the $t(G-1)$ distribution. A better approach, at least in
theory, is to use the studentized bootstrap, or percentile\tkk-$t$,
confidence interval advocated in \citet{Hall_1992}, which is 
\begin{equation}
\label{eq:studboot}
\big[\hat\beta - s_\beta\tk c^*_{\smash{1-\alpha/2}}, \;
\hat\beta - s_\beta\tk c^*_{\smash{\alpha/2}}\big],
\end{equation}
where $s_\beta$ is the standard error of $\hat\beta$ from the CRVE,
and $c^*_z$ denotes the $z$ quantile of the bootstrap
$t$-statistics~$\tau_b^*$. Although the higher-order theory in 
\citet{DMN_2019} does not explicitly deal with confidence intervals, it
strongly suggests that the intervals \eqref{eq:sstarint} and 
\eqref{eq:studboot} should not perform as well as inverting a bootstrap
test based on a bootstrap DGP that imposes the null hypothesis. 
Simulation results in \citet{JGM_2015} are consistent with these 
predictions. However, the intervals \eqref{eq:sstarint} and 
\eqref{eq:studboot} have the advantage that they are easy to compute. 
No iteration is required, and a single set of bootstrap samples can be 
used to compute confidence intervals for all the parameters of interest.

\subsection{Pairs Cluster Bootstrap}
\label{subsec:pairs}

The most important aspect of any bootstrap procedure is how the
bootstrap samples are generated. The only procedure applicable to
every model that uses clustered data is the pairs cluster bootstrap,
which is also sometimes referred to as the cluster bootstrap, the
block bootstrap, or resampling by cluster. The pairs cluster bootstrap
works by grouping the data for every cluster into a $[\biy_g,\biX_g]$
pair and then resampling from the $G$ pairs. Every bootstrap sample
is constructed by choosing $G$ pairs at random with equal 
probability~$1/G$.

Although this procedure ensures that every bootstrap sample contains
$G$ clusters, the number of observations inevitably varies across the
bootstrap samples, unless all cluster sizes are the same. The size of
the bootstrap samples can vary greatly, because the largest clusters
may be over-represented in some bootstrap samples and
under-represented in others. This limits the ability of the bootstrap
samples to mimic the actual sample. So does the fact that the
$\biX^\top\biX$ matrix is different for every bootstrap sample.

Because the pairs cluster bootstrap does not impose the null
hypothesis, care must be taken when calculating the bootstrap test
statistics. If the null hypothesis is that $\beta=\beta_0$, the actual
$t$-statistic will have numerator $\hat\beta-\beta_0$, but the
bootstrap $t$-statistic must have numerator $\hat\beta^*_b-\hat\beta$,
where $\hat\beta^*_b$ is the estimator of $\beta$ for the \th{b}
bootstrap sample. In this case, $\hat\beta$ is the parameter value
associated with the bootstrap DGP. Because the bootstrap DGP does not
impose the null hypothesis, the pairs cluster bootstrap cannot be used
to construct the confidence interval \eqref{eq:confint}, but it can be
used to construct the intervals \eqref{eq:sstarint} and
\eqref{eq:studboot}.

\citet[Section VI]{CM_2015} discusses several problems that can arise
with the pairs cluster bootstrap and sensibly suggests that 
investigators should examine the empirical distributions of the 
bootstrap coefficient estimates and test statistics. For example, if 
the bootstrap distribution has more than one mode, then it probably 
does not provide a good approximation to the actual distribution. This 
can happen when one or two clusters are very different from all the 
others; see \Cref{subsec:leverage}.

\citet{FP_2019} proposes several bootstrap procedures that can be
thought of as variants of the pairs (not pairs cluster) bootstrap. The
first step is to run regressions at either the individual level or the
group $\times$ time\tkk-period level, then aggregate the residuals so
that there is just one residual per cluster, and finally run bootstrap
regressions on the resampled residuals. These procedures include
parametric methods to correct for the heteroskedasticity generated by
variation in the number of observations per group. Remarkably, they
can work well even with just one treated cluster. However, this is
possible only because, unlike methods based on a CRVE, they do not
allow for unrestricted heteroskedasticity.

In general, the pairs cluster bootstrap is expensive to compute.
However, a computational shortcut for linear regression models can
make it feasible even when $N$ and $B$ are both large; see
\citet{JGM-fast}. Nevertheless, we do not recommend this method for
the linear regression model \eqref{eq:lrmodel}, because, as we discuss
in the next subsection, a much better method is available. With
nonlinear models such as the probit model, the pairs cluster bootstrap
may be attractive even though it can be expensive. However, we cannot
recommend it without reservation, because it appears that very little
is known about its finite\tkk-sample properties. Simulation evidence
suggests that, at least in some cases, it can either over-reject or
under-reject severely \citep{MW-TPM,JGM-fast}.

\subsection{Wild Cluster Bootstrap}
\label{subsec:wild}


The restricted wild cluster, or WCR, bootstrap was first suggested in
\citet{CGM_2008}. Until recently, the only variant of it ever used in
practice is based on CV$_{\tn1}$, and, at time of writing, this
classic variant is the only one for which software is readily
available.  Other variants will be discussed briefly at the end of
this subsection.

Suppose that $\tilde\bbeta$ denotes the OLS estimator of $\bbeta$
subject to the restriction $\bia^\top\bbeta=\bia^\top\bbeta_0$, which
is to be tested, and $\tilde\biu_g = \biy_g - \biX_g\tilde\bbeta$
denotes the vector of restricted residuals for the \th{g} cluster.
Then the traditional way to write the WCR bootstrap DGP is
\begin{equation}
\biy_g^{*b} = \biX_g\tilde\bbeta + \biu_g^{*b}, \quad
\biu_g^{*b} = v_g^{*b} \tilde\biu_g, \quad g=1,\ldots,G,
\label{eq:wcr}
\end{equation}
where the $v_g^{\ast b}$ are independent realizations of an auxiliary
random variable $v^*$ with zero mean and unit variance. In practice,
the best choice for $v^*$ is usually the Rademacher distribution, in
which case $v^*$ equals 1 or $-1$ with equal probabilities
\citep{DF_2008,DMN_2019}. This imposes symmetry on the bootstrap 
disturbances.

Instead of using \eqref{eq:wcr}, we can generate the bootstrap scores
directly as
\begin{equation}
\label{eq:scoreboot}
\bis_g^{*b} = v_g^{*b}\tilde\bis_g, \quad g=1,\ldots,G,
\end{equation}
where $\tilde\bis_g = \biX_g^\top \tilde\biu_g$ is the score vector
for the \th{g} cluster evaluated at the restricted estimators. Plugging
the $\bis_g^{*b}$ into \eqref{eq:betahat} then yields the bootstrap
estimators $\hat\bbeta^*_b$. This method is computationally inexpensive
\citep{JGM-fast}. It also provides an intuitive justification for the
WCR bootstrap because, as we stressed in \Cref{sec:asytheory}, the
finite\tkk-sample properties of cluster-robust tests depend mainly on
the properties of the scores.


\citet{DMN_2019} establishes the asymptotic validity of the classic
WCR bootstrap with CV$_{\tn1}$ standard errors and also studies the
unrestricted wild cluster, or WCU, bootstrap. The difference between
the WCR and WCU bootstraps is that, for the latter, the unrestricted
scores $\hat\bis_g$ are used instead of the restricted ones
$\tilde\bis_g$ in the bootstrap DGP~\eqref{eq:scoreboot}. In general,
the WCU bootstrap does not perform as well in finite samples as the
WCR one. The paper contains both theoretical results (based on
Edgeworth expansions) and simulation results to support this
assertion.  However, the WCU bootstrap has the advantage that the
bootstrap DGP does not depend on the restrictions to be tested. The
same set of bootstrap samples can therefore be used to perform tests
on any restriction or set of restrictions and/or to construct
confidence intervals based on either \eqref{eq:sstarint} or
\eqref{eq:studboot} for any coefficient of interest.

\citet{DMN_2019} also proves the asymptotic validity of two variants
of the ordinary wild bootstrap, restricted (WR) and unrestricted (WU),
for the model \eqref{eq:lrmodel}. The ordinary wild bootstrap uses $N$
realizations of $v^*$\tn, one for each observation, instead of just
$G$. This means that the disturbances for the bootstrap samples are
uncorrelated within clusters. Although this implies that the
distribution of the $\hat\bbeta^*_b$ cannot possibly match that of
$\hat\bbeta$, it often does not prevent the distribution of the
$\tau^*_b$ from providing a good approximation to the distribution
of~$\tau$. With one important exception (see below), the WR bootstrap
seems to work less well than the WCR bootstrap, as asymptotic theory
predicts. It can also be much more expensive to compute when $N$ is
large.


The classic versions of the WCR and WCU bootstraps that use
CV$_{\tn1}$ are surprisingly inexpensive to compute. The computations
require only the matrices $\biX_g^\top\biX_g$ and the vectors
$\biX_g^\top\biy_g$; see \citet{JGM-fast}. The calculations can be
made even faster by rewriting the bootstrap test statistic so that it
depends on all the sample data in the same way for every bootstrap
sample; see \citet*{RMNW}. The only thing that varies across the
bootstrap samples is the $G$-vector $\biv^{*b}$ of realizations of the
auxiliary random variable. There are some initial computations that
may be expensive when $N$ is large, but they only have to be done
once. After that, the $\biv^{*b}$ and the results of the initial
computations are used to compute all the bootstrap test statistics.

This fast procedure is implemented in the package \texttt{boottest}
for both \texttt{Stata} and \texttt{Julia}; for details, see
\citet{RMNW}. In \Cref{sec:empirical}, there is an illustration of how
fast it can be; see the notes to \Cref{tab:teenreg}. Importantly,
\texttt{boottest} not only computes WCR bootstrap $P$~values for both
$t$-tests and Wald tests; it also computes WCR bootstrap confidence
intervals based on \eqref{eq:confint}. The package has many other
capabilities as well.

In many cases (exceptions will be discussed below), the classic
CV$_{\tn1}$ variant of the WCR bootstrap yields very accurate
inferences. These are generally more accurate than those for the pairs
cluster or WCU bootstraps. In addition to \citet{DMN_2019}, see
\citet{CGM_2008}, \citet{MW-JAE}, and \citet{JGM-fast}. Because the
$\biX_g$ are always the same, the wild cluster bootstrap is able to
replicate what is often the main source of heterogeneity, namely,
variation in cluster sizes, in every bootstrap sample. This is not the
case for the pairs cluster bootstrap, and it surely contributes to the
superior accuracy of inferences based on the WCR bootstrap.

Of course, no bootstrap method can work perfectly. Not surprisingly,
the performance of the WCR bootstrap using CV$_{\tn1}$ tends to
deteriorate as $G$ becomes smaller, as the number of regressors
increases, and as the clusters become more heterogeneous. In
particular, it can sometimes perform very badly when the number of
treated clusters $G_1$ is very small \citep{MW-JAE, MW-TPM, MW-EJ}.
This is true both for pure treatment models, where all the
observations in each cluster are either treated or not, and for DiD
models, where only some observations in the treated clusters are
treated.


Unlike other methods, which generally over-reject severely when there
are few treated clusters (\Cref{subsubsec:few}), the WCR bootstrap
usually under-rejects in this case. This happens because the
distribution of the bootstrap statistics $\tau^*_b$ depends on the
value of the actual test statistic~$\tau$. The larger is $\tau$, the
more dispersed are the $\tau^*_b$. This makes $\hat P^*(\tau)$ in
\eqref{bootp} larger than it should be. In the most extreme case of
just one treated cluster, rejection frequencies may be essentially
zero. In this case, the bootstrap distribution is often bimodal
\citep[Figure~4]{MW-JAE}, so that plotting it can provide a useful
diagnostic. When there are few treated clusters and the WCR bootstrap
$P$~value seems suspiciously large, it may be worth trying the
ordinary wild restricted (WR) bootstrap, which can sometimes work
surprisingly well in this context \citep{MW-EJ}. This is a case where
randomization inference can be attractive; see \Cref{subsec:RI}.


We recommend using at least one variant of the WCR bootstrap
(preferably with at least $B=9,\tn999$) almost all the time. All
variants are often remarkably inexpensive to compute, and they often
seem to work well. When $G$ is not too small and the clusters are not
too heterogeneous, WCR bootstrap $P$~values and confidence intervals
may be quite similar to ones based on CV$_{\tn1}$ or CV$_{\tn3}$ with
$t(G-1)$ critical values. In that case, it is likely that
finite\tkk-sample issues are not severe, and there is probably no need
to do anything else. When there is a large discrepancy between the
results of different methods, however, it would make sense to try
several bootstrap methods and maybe some of the methods discussed in
\Cref{sec:other}.

The classic version of the WCR bootstrap can sometimes work remarkably
well even when $G$ is very small. In fact, \citet*{CSS_2021} shows
that it can yield exact inferences in certain cases where $N$ is large
and $G$ is small. These results are obtained by exploiting the
relationship between the WCR bootstrap with Rademacher auxiliary
random variables and randomization inference (\Cref{subsec:RI}).
However, they require rather strong homogeneity conditions on the
distribution of the covariates across clusters, as well as limits on
the amount of dependence within each cluster similar to those in
\citet{BCH_2011}.




At this point, a word of warning is in order. Almost all the
simulation results that we have referred to are based on models with
one or just a few regressors, and these regressors are typically
generated in a fairly simple way. Moreover, almost all existing
simulations focus on $t$-statistics rather than Wald statistics. There
is evidence that rejection frequencies for all methods increase as the
number of regressors increases, and that Wald tests are less reliable
than $t$-tests; see \citet[Section~C.2]{DMN_2019} and
\citet{MNW-bootknife} on the former point, \citet{PT_2018} on the
latter, and \citet{JGM-fast} on both of them. Thus the classic WCR
bootstrap may perform less well in empirical applications with large
numbers of regressors than it has typically done in simulations,
especially when there is more than one restriction.



When $G$ is small, the WCR bootstrap encounters an important practical
problem. For the Rademacher distribution, or any other two\tkk-point
distribution, the number of possible bootstrap samples is just
$2^G$\tn. \citet{Webb_2014} proposes a six-point distribution which
largely solves this problem, because $6^G$ is much larger than
$2^G$\tn. This distribution seems to work almost as well as Rademacher
for most values of $G$, and sometimes much better when $G$ is very
small. Whenever either $2^G$ (for Rademacher) or $6^G$ (for six-point)
is smaller than the chosen value of $B$, we can enumerate all possible
bootstrap samples instead of drawing them at random. For example, 
when $G=16$, there are just $65,\tn536$ Rademacher bootstrap samples 
to enumerate (one of which is identical to the actual sample). This
eliminates simulation randomness from the bootstrap procedure. In
fact, \texttt{boottest} uses enumeration by default whenever $B$ is 
greater than the number of possible bootstrap samples.



Most of the discussion above pertains to the classic variant of the
WCR bootstrap, which uses \eqref{eq:scoreboot} to generate bootstrap
samples and CV$_{\tn1}$ to calculate standard errors.
\citet{MNW-bootknife} introduces three new variants. The simplest of
these still uses \eqref{eq:scoreboot}, but the actual and bootstrap
test statistics employ CV$_{\tn3}$ standard errors, computed using
\eqref{eq:jackvar}. Two somewhat more complicated variants generate
the bootstrap scores using an alternative to \eqref{eq:scoreboot} that
implicitly involves a transformation of the restricted scores similar
to the one that yields the $\acute\bis_g$ in \eqref{eq:CV3}. One of
these new variants uses CV$_{\tn1}$ standard errors, and the other
uses CV$_{\tn3}$ standard errors. Simulation results in
\citet{MNW-bootknife} suggest that all three new variants of the WCR
bootstrap outperform the classic variant in a number of circumstances.
In particular, the tendency to over-reject seems to increase much more
slowly for the new variants than for the classic one as the number of
regressors increases and as the variation in cluster sizes increases.
However, the new variants still tend to under-reject severely when the
number of treated clusters is small. We hope that \texttt{Stata} and/or 
\texttt{R} packages able to compute these new variants will become 
available in the near future.




\section{Other Inferential Procedures}
\label{sec:other}



Bootstrap methods are not the only way to obtain inferences more
accurate than those given by cluster-robust $t$-tests and confidence
intervals using the $t(G-1)$ distribution. Numerous other procedures
have been proposed, which broadly fall into two categories that we
discuss in the following two subsections. Except in the case of
treatment models with very few clusters or very few treated clusters
(\Cref{subsec:RI}), we do not recommend any of these procedures in
preference to CV$_{\tn3}$ combined with the $t(G-1)$ distribution or
the WCR bootstrap, especially the newest variants of the latter.
However, when the recommended methods yield conflicting results, it is
surely a good idea to try other methods as well.



\subsection{Alternative Critical Values}
\label{subsec:altcrit}





Critical values for cluster-robust test statistics can be based on
various approximations. The first paper to take this approach seems to be
\citet{BM_2002}. It suggests methods for calculating approximations to
CV$_{\tn2}$ or CV$_{\tn3}$ $t$-statistics based on the Student's $t$
distribution with an estimated degrees-of-freedom (d-o\tkk-f)
parameter. These employ what is called a ``Satter\-thwaite
approximation'' to calculate the d-o\tkk-f. This is done under the
assumption that the variance matrix of $\biu$ is proportional to an
identity matrix. The d-o\tkk-f parameter is different for every
hypothesis to be tested, and it can be much less than~$G-1$.


\citet{Imbens_2016} proposes a similar procedure for $t$-tests based
on CV$_{\tn2}$ under the assumption that the variance matrix of $\biu$ 
corresponds to a random-effects model. As we saw in 
\Cref{subsec:sources}, such a model implies that the disturbances 
within each cluster are equi-correlated, and the intra-cluster
correlation $\rho$ must be estimated from the residuals. When cluster 
fixed effects are partialed out, doing so absorbs any random effects, 
making this approach inapplicable.



\citet{AY-exact} proposes a related method that uses CV$_{\tn1}$
instead of CV$_{\tn2}$. It involves two steps. In the first step, the
CV$_{\tn1}$ standard error for the coefficient of interest is
multiplied by a factor greater than one. In the second step, a
d-o\tkk-f parameter is calculated. The standard error and the
d-o\tkk-f parameter can then be used to compute either a $t$-statistic
and its $P$~value or a confidence interval. A \texttt{Stata} package
called \texttt{edfreg} is available.


The three procedures just discussed are described in some detail in
\citet[Appendix~B]{MW-EJ}. However, the the CV$_2$-based and
CV$_3$-based procedures can be very expensive as described there, and
\citet{NAAMW_2020} provides a better way to compute them. Limited
simulation evidence suggests that the performance of Young's method is
similar to those of the two methods based on CV$_{\tn2}$. However,
this evidence is by no means definitive, because the simulations focus
on only a narrow set of treatment models.

Using Hotelling's $T^{\tk2}$ distribution with estimated degrees of
freedom, \citet{PT_2018} generalizes the CV$_{\tn2}$ procedure of
\citet{BM_2002} to the case of Wald tests based on \eqref{Waldstat}.
Simulations suggest that the resulting tests always reject less often
than standard Wald tests. They rarely over-reject but often
under-reject, and they sometimes do so quite severely. The
\texttt{clubSandwich} package for \texttt{R} and the
\texttt{reg$\_$sandwich} package for \texttt{Stata} implement the
procedures discussed in \citet{PT_2018}.


Although the procedures discussed in this subsection have some
theoretical appeal and seem to work well in many cases, we are not
aware of any evidence that they outperform either tests based directly 
on CV$_{\tn3}$ or the wild cluster bootstrap for a range of models and
DGPs. One limitation is that the Student's $t$ distribution is not very
flexible. Even when the d-o\tkk-f parameter is estimated very well, the
best we can hope for is that a test based on this distribution will be
accurate for some level of interest. It may well over-reject at some
levels and under-reject at others.




Our recommendation is to use the methods discussed in this subsection
primarily to confirm (or perhaps cast doubt on) the results of the
methods that we recommend when there is concern about the reliability
of the latter. Cases of particular concern are ones with few but
balanced clusters (say, $G \leq 12$), ones with balanced but few
treated (or few control) clusters (say $G_1 \leq 6$ or $G-G_1 \leq
6$), ones with seriously unbalanced cluster sizes (even when $G$ is
quite large), ones with treated clusters that are unusually large or
small, and ones with any sort of heterogeneity that causes a few
clusters to have high leverage; see \Cref{subsec:leverage}. It is
always reassuring when several methods yield essentially the same
inferences.





%

\subsection{Randomization Inference}
\label{subsec:RI}

Randomization inference (RI) was proposed by \citet{Fisher_1935} as a 
distribution-free way to perform hypothesis tests in the context of
experiments. RI tests are also called permutation tests.
\citet[Chapter~15]{Lehmann_2005} gives a formal introduction, and
\citet[Chapter~15]{Imbens_2015} provides a more accessible discussion.
The key idea of RI is to compare an outcome $\tau$ that is actually
observed with a set of outcomes that might have been observed if
treatment had been assigned differently. The outcome could be a sample
average, a coefficient estimate, or a test statistic. 

Specifically, consider a clustered regression model with treatment at 
the cluster level, which could be a DiD model where only some 
observations within the treated clusters receive treatment. Then
$\tau$ might be the average treatment effect for some outcome measure.
RI procedures may be attractive when it is plausible that treatment
was assigned at random and the researcher is interested in the sharp
null hypothesis that the treatment has no effect. They can sometimes
yield reliable results even when the number of clusters is small
and/or the number of treated clusters is small.

Suppose there are $G$ clusters, $G_1$ of which received treatment. The
number of ways in which treatment could have been assigned to $G_1$
out of the $G$ clusters is
\begin{equation}
{}_{G}C_{G_1} = \frac{G\tkk!}{G_1!(G-G_1)!}\tk.
\label{Nassign}
\end{equation}
One of these corresponds to the actual assignment, and the remaining
$S = {}_{G}C_{G_1} - 1$ are called re\tkk-randomizations. Each 
re\tkk-randomization involves pretending that a particular set of
$G_1$ clusters was treated, with the remaining $G-G_1$ serving as
controls.  The values of the dependent variable do not change across
re\tkk-randomizations, only the values of the treatment dummy. For
every re\tkk-randomization, indexed by $j$, we could calculate a test 
statistic $\tau^*_j$. If the observable characteristics of the
clusters were all the same, it would make sense to compare $\tau$ with
the empirical distribution of the $\tau^*_j$. To do so, it is natural
to calculate the $P$~value for an upper-tail test as either
\begin{equation}
P_1^*(\tau) = \frac1S \sum_{j=1}^S \IF(\tau^*_j \ge \tau) \quad\mbox{or}
\quad P_2^*(\tau) = \frac1{S+1}\Big(\!1 + \sum_{j=1}^S
\IF(\tau^*_j \ge \tau)\tn\Big).
\label{Pvalues}
\end{equation}
Here $P_2^*$ implicitly includes the actual assignment to treatment,
and $P_1^*$ omits it.

When $\alpha(S+1)$ is an integer for a test at level $\alpha$, both $P$
values in \eqref{Pvalues} yield the same result, and the test is exact
if the distributions of the $\tau^*_j$ are the same as that of~$\tau$.
However, $P_1^*$ and $P_2^*$ can differ noticeably when $S$ is small.
The latter is more conservative and seems to be more popular. For
moderate values of $S$, it is often easy to enumerate all of the
possible $\tau^*_j$, but this is infeasible even when $G$ is not
particularly large. In such cases, we must choose a number of
re\tkk-randomizations, say $B=99,\tn999$, at random. In principle,
these should be drawn without replacement, but that is not important
if $B$ is small relative to~$S$.

It seems to be widely believed that tests based on RI are always 
exact. This is not true. When treatment is not assigned at random, or 
when the observed characteristics of the treated clusters differ 
systematically from those of the controls, we cannot expect the 
distributions of $\tau$ and the $\tau^*_j$ to coincide.


For DiD models, \citet{CT_2011} proposes a test that is very similar
to an RI test based on the OLS estimate of the coefficient on a
treatment dummy. That paper also shows how to obtain a confidence
interval by inverting the test. \citet{MW-RI} studies two procedures
for DiD tests, one based on coefficient estimates (called RI-$\beta$)
and one based on cluster-robust $t$-statistics (called RI-$t$). Both
procedures work very well when the clusters are homogeneous and $G$ is
reasonably large, even when~$G_1=1$. However, when the treated
clusters are systematically larger or smaller than average, neither
RI-$\beta$ nor \mbox{RI-$t$} tests perform well, although the latter
typically perform better. In such cases, it appears that $G$ may have
to be quite large (much larger than for the WCR bootstrap) before
either procedure works really well, even when $G_1$ is not
particularly small.

\citet{AY_fisher} contains an interesting recent application of RI,
where the RI-$\beta$ and \mbox{RI-$t$} procedures are applied to 
regressions for the results of 53 randomized experiments in published 
papers. \texttt{Stata} packages called \texttt{randcmd} and 
\texttt{randcmdci} are available. In many cases, estimates that were 
originally reported to be significant are not significant according to 
the RI procedures. Not surprisingly, this is particularly true for 
regressions with a few high-leverage clusters or observations.




There is evidently a close relationship between RI and the wild cluster
bootstrap. Evaluating all possible bootstrap samples by enumeration is 
quite similar to evaluating all possible re\tkk-randomizations. The 
results of \citet{CSS_2021} and \citet{MW-RI} strongly suggest that 
homogeneity across clusters is more important for RI than for the WCR 
bootstrap. For the former, the regressors change as we compute each of 
the $\tau^*_j$, but the regressand stays the same. For the latter, the 
regressand changes as we compute each of the $\tau^*_b$, but the 
regressors stay the same. Thus, the empirical distribution to which 
$\tau$ is being compared is conditional on the actual regressors 
(including the clusters actually treated) for the WCR bootstrap, but 
not for the RI procedures discussed above.

The RI methods discussed so far are not the only ones.
\citet*{CRS_2017} proposes ``approximate randomization tests'' based
on the cluster-level estimators of \citet{IM_2010}; see 
\citet*{CCKS_2021} for a guide to this approach. In models for
treatment effects, every cluster must include both treated and
untreated observations. This can be accomplished by merging clusters,
but at the cost of making $G$ smaller with a resulting loss in power.
\citet{Hag_2019a,Hag_2019b} develops RI tests that can be used even
when $G$ is quite small and there is substantial heterogeneity across
clusters. These tests do not require cluster-level estimation, but
$G_1$ and $G-G_1$ should both be no less than~4. \citet{HS_2020}
develops an RI procedure based on RI-$t$ for the case in which one
cluster is much larger than any of the others.

An alternative way to deal with the problem of one or very few treated
clusters, which can involve an RI-like procedure, is the method of
synthetic controls surveyed in \citet{Abadie_2021}.

\section{What Should Investigators Report?}
\label{sec:report}







Many studies fail to report enough information to convince readers
that their empirical results should be believed. Investigators often
simply report test statistics or confidence intervals based on
CV$_{\tn1}$ together with the $t(G-1)$ distribution. Although results
of this type may be reliable, they often will not be. Unless the
number of clusters is fairly large and evidence demonstrates a
convincing degree of homogeneity across clusters, results from at
least one or two other methods should be reported. These might include
tests and/or confidence intervals based on CV$_{\tn3}$, $P$ values
and/or confidence intervals based on one or more variants of the WCR
bootstrap discussed in \Cref{subsec:wild}, or perhaps results from
some of the alternative methods discussed in \Cref{sec:other}.


It is important to report some key information about the sample as a
matter of routine. The fundamental unit of inference when the
observations are clustered is not the observation but the cluster;
this is evident from \eqref{eq:betahat} and \eqref{eq:trueV}. The
asymptotic theory discussed in \Cref{subsec:large} therefore depends
on $G$, not~$N$\tn. With the exception of certain very special cases
discussed in \Cref{subsec:small}, asymptotic approximations tend to
work poorly when there are few clusters. It is therefore absolutely
essential to report the number of clusters, $G$, whenever inference is
based on a~CRVE. This is even more important than reporting~$N$\tn.

Moreover, because the distributional approximations perform best when
the scores are homogeneous across clusters, and the most important
source of heterogeneity is often variation in cluster sizes, it is
extremely important to report measures of this variation. At a
minimum, we believe that investigators should always report the median
cluster size and the maximum cluster size, in addition to $N$ and $G$.
When $G$ is small, or when the distribution of the $N_g$ is unusual,
it would be good to report the entire distribution of cluster sizes in
the form of either a histogram or a table.


Of course, clusters can be heterogeneous in many ways beyond their
sizes. In \Cref{subsec:leverage}, we discussed a measure of leverage
at the cluster level. Another measure is the partial leverage of each
cluster for the coefficient(s) of interest, which generalizes the
observation-level notion of partial leverage of \citet{CW_1980}. If
$\acute\bix_j$ denotes the vector of residuals from regressing
$\bix_j$ on all the other regressors, and $\acute\bix_{gj}$ is the
subvector of $\acute\bix_j$ corresponding to the \th{g} cluster, then
the partial leverage of cluster $g$ for the \th{j} coefficient is
\begin{equation}
\label{partlev}
L_{gj} =
\frac{\acute\bix_{gj}^\top\acute\bix_{gj}}{\acute\bix_j^\top\acute\bix_j}\tk.
\end{equation}
When a cluster has high partial leverage for the \th{j} coefficient,
removing that cluster has the potential to change the estimate of the
\th{j} coefficient substantially.


A popular way to quantify the heterogeneity of clusters is to
calculate $G_j^*$\tn, the ``effective number of clusters'' for the
\th{j} coefficient, as proposed in \citet*{CSS_2017}. This number is
always less than $G$, and it can provide a useful warning when $G_j^*$
is much smaller than~$G$. The value of $G_j^*$ depends on an unknown
parameter $\rho$, the intra-cluster correlation of the disturbances
when they are assumed to be equi-correlated. \citet{CSS_2017} suggests
setting $\rho=1$, as a sort of worst case. However, since cluster
fixed effects absorb all intra-cluster correlation for the random
effects model \eqref{eq:REmodel}, only $\rho=0$ makes sense for models
with such fixed effects, and trying to use $\rho\ne0$ can lead to
numerical instabilities. \citet{MNW-influence} discusses these issues
and shows how to calculate $G_j^*(\rho)$ efficiently for any value of
$\rho$ in models without cluster fixed effects.


Using \eqref{partlev} and the definition of $G_j^*(0)$, it can be
shown that the latter is a monotonically decreasing function of the
squared coefficient of variation of the partial leverages $L_{gj}$, 
which is defined as
\begin{equation*}
V_s(L_{j\bullet}) = \frac{1}{G\tk \bar L_j^2}
\sum_{g=1}^G (L_{gj} - \bar L_j)^2,
\end{equation*}
where $\bar L_j$ is the sample mean of the $L_{gj}$. Thus $G_j^*(0)$
and $V_s(L_{j\bullet})$ provide exactly the same information. When the
partial leverage for the \th{j} coefficient is the same for every
cluster, $V_s(L_{j\bullet}) = 0$ and $G_j^*(0) = G$. When the $L_{gj}$
differ greatly, $V_s(L_{j\bullet})$ is large, and $G_j^*(0) <\tn< G$.

We suggest that investigators should routinely report the leverages
$L_g$ defined in \eqref{eq:traceXX}, the partial leverages $L_{gj}$
for the coefficient(s) of interest, and the $\hat\beta_j^{(g)}$ for
the same coefficient(s), at least when they provide information beyond
that in the distribution of the cluster sizes. When $G$ is small, it
may be feasible to report all these numbers. Otherwise, it may be
feasible to graph them, as in \Cref{fig:EDF}, or to report summary
statistics such as $G_j^*(0)$ or $V_s(L_{j\bullet})$. Reporting
measures of influence and leverage should help to identify cases in
which inference may be unreliable, as well as sometimes turning up
interesting, or possibly disturbing, features of the data.

When there is clustering in two or more dimensions
(\Cref{subsec:twoway}), we recommend that investigators compute
everything just suggested for each of the clustering dimensions and
also for their intersection(s). If the results are interesting, they
should be reported. For two\tkk-way clustering, this might mean
reporting, or at least summarizing, up to three sets of leverages,
partial leverages, and $\hat\beta_j^{(g)}$\tn.


\section{Empirical Example}
\label{sec:empirical}



We illustrate many of our recommendations by revisiting a
long-standing empirical question in labor economics, namely, the
impact of the minimum wage on young people. In the past two decades,
many U.S.\ states have significantly increased their minimum wages. In
fact, from 2000 to 2019, every state increased the nominal minimum
wage by at least 27\%, and six states doubled it. Moreover, recent
proposals to increase the national minimum wage to \$15 per hour have
reinvigorated the debate about the effects of minimum wages.

Some classic references on the impact of the minimum wage are
\citet{Mincer_1976} and \citet{CK_1994}. The latter paper was among
the very first and most influential applications of the DiD
methodology, which continues to be used in this literature. For
example, \citet{Seattle_2017} uses a DiD analysis to study the effects
of a large increase in the minimum wage in Seattle.
\cite{Wolfson_2019} and \cite{Neumark_2021} survey many recent studies
on the impacts of minimum wages. Both conclude that the majority of
studies, but not all, find dis-employment effects that are
concentrated among teenagers and those with low levels of education.
\citet{Manning_2021} explores why the evidence on employment effects
is mixed.

Instead of using a DiD approach, we exploit state\tkk-level
differences in the minimum wage and analyze their impacts on
labor-market and education outcomes at the individual level. Although
we treat the minimum wage as exogenous in our analysis, we hesitate to
call our estimates causal. There is reason to believe that
state\tkk-level minimum wages may be endogenous, because states may be
more likely to increase them during good economic times. Moreover, it
is possible that the effects of the minimum wage differ depending on
the state of the economy. However, we ignore these issues, because our
principal objective is to illustrate the importance of clustering for
statistical inference.
 
The model we estimate is
\begin{equation}
\label{reg:wage}
y_{ist} = \alpha + \beta\tk \textrm{mw}_{\tn st} + \biZ_{ist}\bgamma
+ \textrm{year}_t\tk\bdelta_t + \textrm{state}_s\tk\bdelta_s
+ u_{ist}.
\end{equation}
Here $y_{ist}$ is the outcome of interest for person $i$ in state $s$
in year~$t$. There are three outcome variables. ``Hours'' records the
usual hours worked per week, which is defined only for employed
individuals. ``Employed'' is a binary variable equal to~1 if person
$i$ is employed and to~0 if they are either unemployed or not in the
labor force. ``Student'' is a binary variable equal to~1 if person $i$
is currently enrolled in school and to~0 otherwise. The parameter of
interest is $\beta$, which is the coefficient on $\textrm{mw}_{\tn
st}$, the minimum wage in state $s$ at time~$t$. The row vector
$\biZ_{ist}$ collects a large set of individual-level controls,
including race, gender, age, and education. There are also year and
state fixed effects. \cite{NW_2007} estimates models similar to
\eqref{reg:wage} with individual-level data, clustering at the state
level.

Data at the individual level from the American Community Survey (ACS)
were obtained from IPUMS \citep{IPUMS_2020} and cover the years
2005--2019. The minimum wage data were provided by
\cite{Neumark_2019}, and we have collapsed them to state\tkk-year
averages to match the ACS frequency. Following previous literature, we
restrict attention to teenagers aged 16--19. We keep only individuals
who are ``children'' of the respondent to the survey and who have
never been married. We drop individuals who had completed one year of
college by age~16 and those reporting in excess of 60~hours usually
worked per week. We also restrict attention to individuals who
identify as either black or white.



We consider six different clustering structures that lead to nine
different standard errors for $\hat\beta$. These are no clustering
(with HC$_1$ standard errors), one\tkk-way clustering at either the
state\tkk-year, state, or region\footnote{Regions are defined as the
U.S.\ Census Divisions, with the following partitioning of states. New
England: CT, MA, MN, NH, RI, VT; Middle Atlantic: NJ, NY, PA; South
Atlantic: DC, DE, FL, GA, MD, NC, SC, VA, WV; East South Central: AL,
KY, MS, TN; East North Central: IL, IN, MI, OH, WI; West North
Central: IA, KS, MN, MO, ND, NE, SD; West South Central: AR, LA, OK,
TX; Mountain: AZ, CO, ID, MT, NM, NV, UT, WY; Pacific: AK, CA, HI, OR,
WA.} level (with both CV$_{\tn1}$ and CV$_{\tn3}$ standard errors),
and two\tkk-way clustering by state and year or by region and year
(with CV$_{\tn1}$ standard errors, the only ones currently available
for multi-way clustering). Early empirical research on the impacts of
the minimum wage would have used either conventional or HC$_1$
standard errors, but modern studies would almost always cluster at
some level.


Since the minimum wage is invariant within state\tkk-year clusters, it
seems highly likely that the scores are correlated within them, and we
therefore consider (at least) state\tkk-year clustering. However,
after a state has increased its minimum wage, it almost always remains
at the new level until it is increased again. This implies that
minimum wages must be correlated across years within each state.
Unless the disturbances happen to be uncorrelated across years within
states, the scores will therefore be correlated, which suggests that
state\tkk-level clustering may be appropriate. We consider
region-level clustering based on the nine census divisions because
there may be correlations among nearby states. Largely for
completeness, we also consider two\tkk-way clustering by either state
or region and year.










\begin{table}[tp]
\caption{Estimated Impact of the Minimum Wage}
\label{tab:teenreg}
\vspace*{-0.5em}
\begin{tabular*}{\textwidth}{@{\extracolsep{\fill}}
lld{2.4}d{2.4}d{2.4}}
\toprule
\multicolumn{1}{l}{Clustering level} & & \multicolumn{1}{c}{Hours} & 
\multicolumn{1}{c}{Employed} & \multicolumn{1}{c}{Student} \\		
\midrule
& $\hat \beta$   & -0.1539 & -0.0037 & 0.0022 \\
\vspace{-11pt}\\
None: HC$_1$ & $t$-statistic & -5.4469 & -5.2801 & 4.9719 \\
& $P$~value, $N(0,1)$ & 0.0000 & 0.0000 & 0.0000 \\
\vspace{-11pt}\\
State\tkk-year: CV$_{\tn1}$ & $t$-statistic & -3.3823 & -2.6492 & 4.0776 \\
& $P$~value, $t(764)$ & 0.0008 & 0.0082 & 0.0001 \\
& $P$~value, WCR & 0.0027 & 0.0145 & 0.0005 \\
\vspace{-11pt}\\
State\tkk-year: CV$_{\tn3}$ & $t$-statistic & -2.9466 & -2.3199 & 3.6095 \\
& $P$~value, $t(764)$ & 0.0033 & 0.0206 & 0.0003 \\
\vspace{-11pt}\\
State: CV$_{\tn1}$ & $t$-statistic & -2.4696 & -1.3679 & 2.9780 \\
& $P$~value, $t(50)$ & 0.0170 & 0.1775 & 0.0045 \\
& $P$~value, WCR & 0.0362 & 0.2141 & 0.0238 \\
\vspace{-11pt}\\
State: CV$_{\tn3}$ & $t$-statistic &-2.2925 &-1.2203 &2.7884 \\
& $P$~value, $t(50)$ &0.0261 &0.2281 &0.0075 \\
\vspace{-11pt}\\
Region: CV$_{\tn1}$ & $t$-statistic & -2.2478 & -1.0230 & 3.1743 \\
& $P$~value, $t(8)$ & 0.0548 & 0.3362 & 0.0131 \\
& $P$~value, WCR & 0.0527 & 0.3826 & 0.0430 \\
\vspace{-11pt}\\
Region: CV$_{\tn3}$ & $t$-statistic &-2.0407 &-0.8827 &2.6715 \\
& $P$~value, $t(8)$ &0.0756 &0.4031 &0.0283 \\
\vspace{-11pt}\\
State \& year: two\tkk-way CV$_{\tn1}$ & $t$-statistic & -2.5197 & -1.4776 & 3.4443 \\
& $P$~value, $t(14)$ & 0.0245 & 0.1617 & 0.0039 \\
& $P$~value, WCR (year) & 0.1281 & 0.2148 & 0.0034 \\
\vspace{-11pt}\\
Region \& year: two\tkk-way CV$_{\tn1}$ & $t$-statistic & -2.2842 & -1.0999 & 3.5766 \\
& $P$~value, $t(8)$ & 0.0517 & 0.3034 & 0.0072 \\
& $P$~value, WCR (region) & 0.0736 & 0.3711 & 0.0367 \\
\bottomrule
\end{tabular*}
\vskip 6pt {\footnotesize \textbf{Notes:} There are 765 state\tkk-year
clusters, 51 state clusters, and 9 region clusters, with
$492{,\tkk}827$ observations in the hours dataset and
$1{,\tkk}531{,\tkk}360$ observations in the employment/student
dataset. The bootstrap dimension for two\tkk-way clustering is given
in parentheses. Classic WCR bootstrap $P$~values employ $B=99,999$,
mostly using the Rademacher distribution. When bootstrapping by
region, they are calculated using Webb's six-point distribution. When
bootstrapping by year, they are calculated by enumerating the
Rademacher distribution, so that $B=32{,}768$. Obtaining the
CV$_{\tn1}$ and WCR bootstrap results in this table took 6 minutes and
12 seconds using \texttt{Stata}~16 and \texttt{boottest}~2.5.3 on one 
core of an Intel~i9 processor running at 3.6~GHz.}
\end{table}



\Cref{tab:teenreg} presents the estimates of $\beta$ from
regression~\eqref{reg:wage} for all three regressands, along with
$t$-statistics and $P$~values for each of the clustering levels
considered. The coefficients are negative for the hours and employment
regressands and positive for the student one. Under the surely false
assumption of no clustering, all of these coefficients are extremely
significant, with $P$ values below $10^{-6}$. We do not compute
bootstrap $P$ values, because it would be prohibitively costly, and
they must all be very close to zero.

When we instead cluster at the state\tkk-year level, the
$t$-statistics become smaller, especially for the employment model.
Nevertheless, clustering at this level still leads us to conclude that
all three coefficients are significant, with WCR bootstrap $P$ values
ranging from 0.0005 to 0.0145. However, some of our conclusions change
radically when we cluster at the state or region levels. For all three
outcome variables, the $t$-statistics become smaller, and the $P$
values (especially the CV$_{\tn3}$ and bootstrap ones) become larger.
The bootstrap $P$ values often differ substantially from the ones
based on the $t(G-1)$ distribution, which is expected given that the
clusters are either very unbalanced or small in number
(\Cref{subsec:emplever}). Clustering by region always yields larger
bootstrap $P$ values than clustering by state.


If instead we adopt the conservative approach of clustering at the 
level with the largest standard error, we cluster at the state level 
for the student model, but at the region level for the hours and 
employment models. Because the number of regions is so small, however,
the latter results may not be reliable, even when we use CV$_{\tn3}$
or the WCR bootstrap. In our view, it is probably most reasonable to
cluster at the state level in all three cases.


Luckily, the choice between state and region clustering does not
matter much. In either case, we find from \Cref{tab:teenreg} that an
increase in the minimum wage is associated with a decrease in
employment, but the coefficient is not significant at any conventional
level using either state or region clustering. It is also associated
with an increased probability of being a student, which is significant
at the 5\% level regardless of clustering level. For hours worked, the
coefficient is negative. It is significant at the 5\% level for state
clustering, but not for region clustering.

The table also reports asymptotic and bootstrap $t$-statistics and $P$
values for two\tkk-way clustering, either by state and year or by
region and year. The bootstrap method we use combines the one\tkk-way
WCR bootstrap with the two\tkk-way variance matrix \eqref{eq:betavar}.
This can be done in two different ways, corresponding to each of the
two clustering dimensions. Based on simulation evidence in
\citet{MNW_2021}, we bootstrap by the dimension with the smallest
number of clusters. Since two\tkk-way clustering does not change most
of the results very much, however, there appears to be little reason
to use it in this case.

We tentatively conclude that an increase in the minimum wage is
associated with a significant decrease in hours worked and
a significant increase in the likelihood of being a student.
\citet{Seattle_2017} find a similar reduction in hours following a
minimum wage increase in Seattle, and the effects of the minimum wage
on school enrollment are discussed in \citet{NW-1995}. For employment,
we obtain a small negative effect, but it is not close to being
statistically significant when we cluster at either the state or
region levels. Thus our results are consistent with, and add support
for, the conclusions of \citet{Manning_2021} about the ``elusive''
employment effect of the minimum wage.

\begin{table}[tp]
\caption{Score\tk-\tn Variance Tests for Level of Clustering}
\label{tab:teenscore}
\vspace*{-0.5em}
\begin{tabular*}{\textwidth}{@{\extracolsep{\fill}}
ld{2.4}d{2.4}d{2.4}d{2.4}d{2.4}d{2.4}}
\toprule
& \multicolumn{1}{c}{Hours} & \multicolumn{1}{c}{Employed} & 
\multicolumn{1}{c}{Student} & \multicolumn{1}{c}{Hours} & 
\multicolumn{1}{c}{Employed} & \multicolumn{1}{c}{Student} \\
\midrule
& \multicolumn{3}{c}{H$_{\rm none}$ vs.\ H$_{\text{state\tkk-year}}$} & 
\multicolumn{3}{c}{H$_{\text{state\tkk-year}}$ vs.\ H$_{\rm state}$}   \\
\cmidrule{2-4}\cmidrule{5-7}
$\hat \tau$ statistic & 10.1108 & 18.2723 & 2.8558 & 2.2800 & 7.9453 & 3.1105 \\
$P$~value, asymptotic & 0.0000 & 0.0000 & 0.0021 & 0.0113 & 0.0000 & 0.0009 \\
$P$~value, bootstrap & 0.0000 & 0.0000 & 0.0026 & 0.0227 & 0.0000 & 0.0085 \\
\midrule
& \multicolumn{3}{c}{H$_{\rm none}$ vs.\ H$_{\rm state}$} &
\multicolumn{3}{c}{H$_{\text{state\tkk-year}}$ vs.\ H$_{\rm region}$}  \\
\cmidrule{2-4}\cmidrule{5-7}
$\hat \tau$ statistic & 10.9183 & 38.9394 & 4.6608 & 2.5226 & 11.2952 & 1.2701  \\
$P$~value, asymptotic & 0.0000 & 0.0000 & 0.0000  & 0.0058 & 0.0000 & 0.1020 \\
$P$~value, bootstrap & 0.0000 & 0.0000 & 0.0007 & 0.0279 & 0.0000 & 0.1388 \\
\midrule
& \multicolumn{3}{c}{H$_{\rm none}$ vs.\ H$_{\rm region}$} & 
\multicolumn{3}{c}{H$_{\rm state}$ vs.\ H$_{\rm region}$} \\
\cmidrule{2-4}\cmidrule{5-7}
$\hat \tau$ statistic & 9.4096 & 49.6559 & 2.7201 & 0.6757 & 2.2566 & -0.3206 \\
$P$~value, asymptotic & 0.0000 & 0.0000 & 0.0033 & 0.2496 & 0.0120 & 0.6257 \\
$P$~value, bootstrap & 0.0000 & 0.0120 & 0.0000 & 0.2662 & 0.0410 & 0.5497 \\
\bottomrule
\end{tabular*}
\vskip 6pt
{\footnotesize \textbf{Notes:} There are 765 state\tkk-year clusters, 51 
state clusters, and 9 region clusters. The test statistic $\hat\tau$
is asymptotically distributed as $\N(0,1)$. All $P$ values are for
one\tkk-sided tests. The bootstrap $P$~values were 
calculated with $B=99,999$.}
\end{table}


To confirm the choice of clustering level, we use the 
score\tkk-variance tests of \citet{MNW-testing} to test for the 
appropriate level of clustering.\footnote{We do not calculate the
tests proposed in \citet{IM_2016}, because \eqref{reg:wage} cannot be 
estimated on a cluster-by-cluster basis for state\tkk-level or 
state\tkk-year clusters without dropping some regressors.}
\Cref{tab:teenscore} presents results from six tests for each of the
three models. Following a systematic, sequential testing approach, we
would test independence vs.\ state\tkk-year, then state\tkk-year vs.\
state, and finally state vs.\ region. Apart from the possibility of
two\tkk-way clustering, we would conclude that the appropriate level
of clustering is that of the first non-rejected null hypothesis. For
the hours model, the score\tkk-variance tests marginally favor region
clustering over state clustering, but, for the employment and student
models, they clearly favor state clustering. Thus, these tests lend
support to our decision to cluster at the state level; see the
discussion of \Cref{tab:teenreg}.

\subsection{Cluster Heterogeneity}
\label{subsec:emplever}





\begin{table}[tp]
\caption{Summary Statistics for Cluster Heterogeneity}
\label{tab:clustsum}
\vspace*{-0.5em}
\begin{tabular*}{\textwidth}{@{\extracolsep{\fill}}
ld{4.0}d{4.2}d{7.0}d{6.0}d{6.0}d{6.0}d{6.0}d{6.0}}
\toprule
Clustering & \multicolumn{1}{c}{$G$} & \multicolumn{1}{c}{$G_\beta^*(0)$}
& \multicolumn{1}{c}{$\bar{N}_g$} & \multicolumn{1}{c}{min.} & 
\multicolumn{1}{c}{$1^{\rm st}$ quart.} & \multicolumn{1}{c}{median} & 
\multicolumn{1}{c}{$3^{\rm rd}$ quart.} & \multicolumn{1}{c}{max.} \\
\midrule
\multicolumn{8}{l}{Hours data: $N = 492{,}827$} \\
\midrule
State\tkk-year &765 &79.4 &644 &6  &176  &480  &860  &3{,}052 \\
State &51 &16.2 &9{,}663 &258 &2{,}495 &7{,}082 &13{,}481 &35{,}995 \\
Year &15 &6.6 &32{,}855 &28{,}262 &28{,}839 &30{,}733 &40{,}224 &40{,}394 \\
Region &9 &7.5 &54{,}759 &27{,}849 &37{,}396 &50{,}489 &65{,}389 &96{,}337 \\
\midrule
\multicolumn{8}{l}{Employment and student data: $N = 1{,}531{,}360$}  \\
\midrule
State\tkk-year &765 &66.0 &2{,}002 &42 &524 &1{,}413 &2{,}426 &10{,}794 \\
State &51 &13.1 &30{,}027 &927 &7{,}363 &22{,}845 &37{,}020 &144{,}914 \\
Year &15 &6.5 &102{,}091
&92{,}701 &95{,}589 &102{,}319 &108{,}858 &110{,}528 \\
Region &9 &7.0 &170{,}151
&74{,}172 &104{,}703 &181{,}767 &208{,}099 &291{,}955 \\
\bottomrule
\end{tabular*}
\vskip 6pt
{\footnotesize \textbf{Notes:} The values of $G_\beta^*(0)$
are calculated using 28 regressors after the state dummies have been
partialed out. The $\beta$ subscript emphasizes the fact that they 
correspond to the coefficient $\beta$ in \eqref{reg:wage}. Because
there are state fixed effects, values of $G_\beta^*(1)$ are not
reported; see \citet{MNW-influence}.}
\end{table}






As we discussed in \Cref{sec:report}, investigators should be
suspicious of any results that are overly dependent on very few
clusters. \Cref{tab:clustsum} reports several summary statistics for
cluster heterogeneity. Specifically, it reports $G$, the number of
clusters and $G_\beta^*(0)$, the effective number of clusters
(\Cref{sec:report}), as well as the average, minimum, first and third
quartiles, median, and maximum of the $N_g$. These statistics suggest
that the state\tkk-year clusters are extremely unbalanced. Although
there are $G=765$ clusters, the effective number $G_\beta^*(0)$ is
smaller than $G$ by a factor of roughly 10 to 12. The maximum cluster
size is over six times the median, and the third quartile is nearly
twice the median.





From \Cref{tab:clustsum}, we also see that the state clusters are
extremely unbalanced, based both on their sample sizes and on the
values of $G_\beta^*(0)$. The region clusters are fairly balanced in
terms of their sample sizes, with the maximum $N_g$ only about four
times the minimum. The values of $G_\beta^*(0)$ are also not
worrisome. The year clusters (which we only use in two\tkk-way
clustering) are very balanced in terms of their sample sizes but much
less so in terms of $G^*(0)$. We would expect asymptotic tests based
on CV$_{\tn1}$ to perform rather poorly in all cases, because there is
a lot of cluster heterogeneity for clustering by state\tkk-year and
state, a small number of clusters for clustering by region, and a
small number of effective clusters for clustering by year.





For each of the three levels of one\tk-way clustering, we calculate
measures of leverage, partial leverage, and influence using the
\texttt{summclust} package in \texttt{Stata} \citep{MNW-influence}.
However, we only report results for state\tkk-level clustering. Since
state\tkk-year clustering is strongly rejected against state\tkk-level
clustering (\Cref{tab:teenscore}), it does not seem interesting to
discuss it. We also do not discuss region-level clustering, because
there are just nine regions, and there is not much evidence in favor
of clustering at that level. 



\begin{figure}[tb]
\begin{center}
\caption{\label{fig:EDF}{Empirical Distribution Functions for 51 State
Clusters}}
{\includegraphics[width=1.00\textwidth]{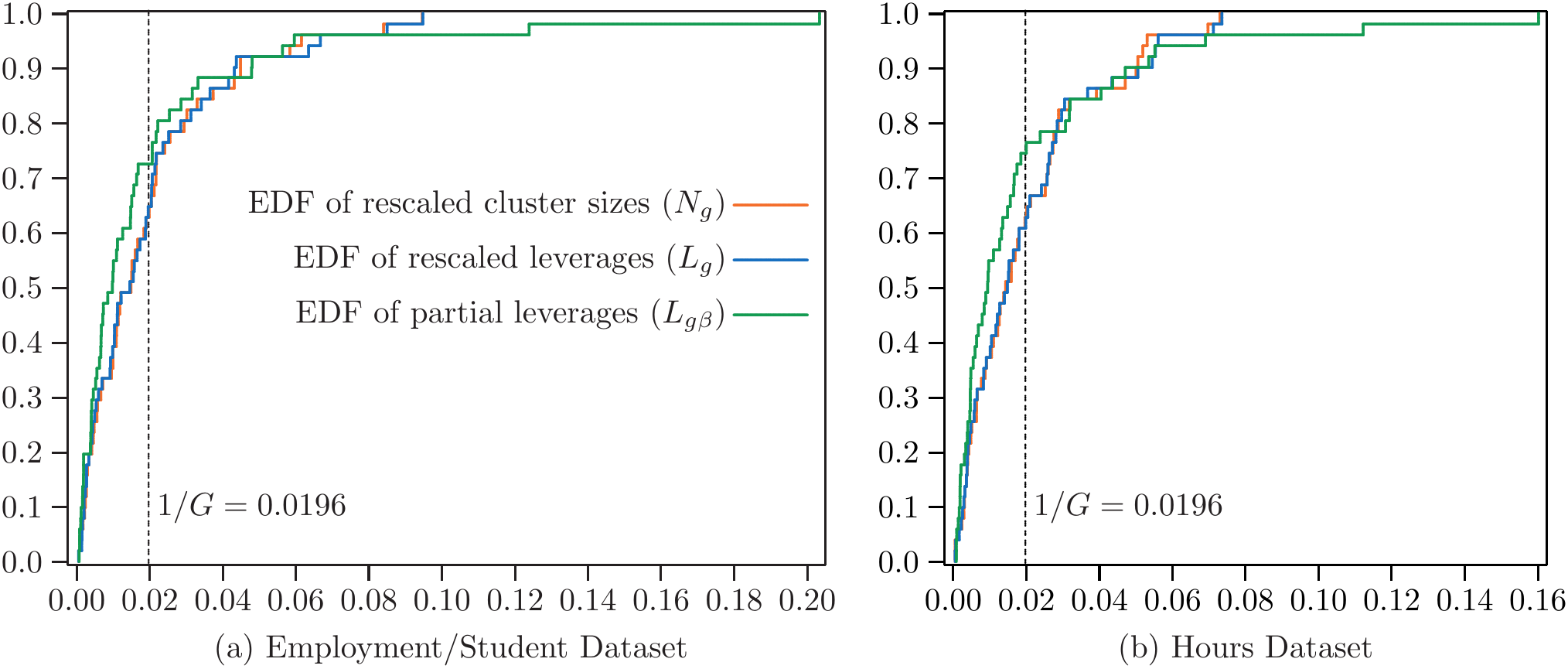}}
\end{center}
\textbf{Notes:} In both panels, each of the three staircase curves is the EDF
for the $N_g$, the $L_g$, or the $L_{g\beta}$. They were plotted in
that order. When they overlap, only the last one to be plotted is
visible. The $N_g$ and the $L_g$ have been rescaled so that they sum
to unity, like the $L_{g\beta}$.
\end{figure}


The empirical distributions of the $N_g$, the $L_g$, and the
$L_{g\beta}$ (the partial leverages) are shown in \Cref{fig:EDF},
where the $N_g$ and the $L_g$, which actually sum to $N$ and $k$,
respectively, have been rescaled so that they sum to unity. In a
perfectly balanced case, where all of the $N_g$ (or $L_g$ or
$L_{g\beta}$) are identical, these EDFs would simply be vertical lines
at~$1/G$. Both datasets are evidently far from being perfectly
balanced. The $L_g$ are essentially proportional to the $N_g$ and
extremely highly correlated with them; the correlations are 0.9986 for
the employment/student dataset and 0.9982 for the hours dataset. This
is evident in both panels, where the EDFs for the $N_g$ and the $L_g$
are extremely similar. Thus, for leverage, it appears that there is no
heterogeneity in the clusters other than what is implied by the
cluster sizes.

In contrast, the partial leverages vary much more across states than
do the other two measures. The largest values of the $L_{g\beta}$ are
considerably larger than the largest values of the $N_g$ and the
$L_g$, especially for the employment/student dataset; note that all of
these are for California. Not coincidentally, more of the $L_{g\beta}$
than of the other measures are smaller than $1/G$, the average. The
correlations between the $L_{g\beta}$ and the $L_g$ are 0.9150 for the
student/employment dataset and 0.9165 for the hours dataset. The
correlations between the $L_{g\beta}$ and the $N_g$ are also very
similar for the two datasets, at 0.9209 and 0.9163.


The substantial amount of cluster heterogeneity that is evident in
both \Cref{tab:clustsum} and \Cref{fig:EDF} suggests that inference
based on CV$_{\tn1}$ and the $t(50)$ distribution may not be reliable.
Further evidence on this point will be provided in
\Cref{subsec:emp-placebo}.



For the student model, we find no evidence of influential clusters.
The $\hat \beta^{(g)}$ range from 0.00185 to 0.00246, with $\hat \beta
= 0.00221$. On the other hand, for the employment model, there are two
noticeable values of~$\hat\beta^{(g)}$: Texas has $\hat\beta^{(g)} =
-0.00529$ and New York has $\hat\beta^{(g)} = -0.00203$, with $\hat\beta
= -0.00367$ and all the remaining $\hat\beta^{(g)}$ in the interval
$[-0.00455 ; -0.00302]$. For comparison, the standard error of
$\hat\beta$ is $0.00268$.

For the hours model, four values of the $\hat\beta^{(g)}$ stand out.
They are Minnesota ($-0.1776$) and Texas ($-0.1770$) at one end
and Arizona ($-0.1274$) and New York ($-0.1227$) at the other. For
comparison, $\hat\beta = -0.1539$, and all the other $\hat\beta^{(g)}$
lie in the interval $[-0.1713; -0.1406]$. In this case, the standard
error of $\hat\beta$ is $0.0623$.

\subsection{Placebo Regressions for the Empirical Example}
\label{subsec:emp-placebo}

As we discussed in \Cref{subsec:placebo}, placebo regressions provide
a simple way to check the level at which the residuals are clustered,
even when the pattern of intra-cluster correlation is unknown and
perhaps very complicated. If a placebo regressor is clustered at, say,
the state level, then the empirical scores will also be clustered at
that level unless the residuals are clustered only at a finer level.
When the residuals display intra-cluster correlation at the state
level, we would therefore expect placebo\tkk-regression tests with
standard errors clustered at that level to reject roughly as often as
they should. But we would expect placebo regressions with standard
errors clustered at finer levels to over-reject.

We perform two sets of placebo\tkk-regression experiments for each of
the three equations estimated in \Cref{tab:teenreg}. In the first set,
the placebo regressor is a DiD-style treatment dummy similar to the
ones used in \citet{BDM_2004} and the other papers cited in
\Cref{subsec:placebo}. Treatment is randomly applied to 15, 20, 25,
30, or 35 states. For each state, it begins randomly in any year
excluding 2005 (to avoid collinearity with the state fixed effects)
and continues through 2019.\footnote{Of course, this sort of DiD model
with two\tkk-way fixed effects is somewhat obsolete; see
\cite{CS_2021} and other papers cited therein. But we would still
expect to find no effects for placebo treatments beyond those
attributable to chance.} Rejection percentages are shown in the top
panel of \Cref{tab:placebo}. The first number in each pair is the
smallest rejection percentage observed over the five experiments for
each equation, and the second number is the largest one. The numbers
of treated and control states are deliberately chosen not to be
extreme, so as to avoid the issues discussed in \Cref{subsubsec:few}.


\begin{table}[tp]
\caption{Rejection Percentages for Placebo Regressions}
\label{tab:placebo}
\vspace*{-0.5em}
\begin{tabular*}{\textwidth}{@{\extracolsep{\fill}}
lcd{2.1}d{2.1}cd{2.1}d{2.1}cd{2.1}d{2.1}cd{2.1}d{2.1}cd{2.1}d{2.1}cd{2.1}d{2.1}}
\toprule
 && \multicolumn{8}{c}{DiD-type treatment} && 
\multicolumn{8}{c}{State\tkk-level AR(1) component}\\
\cmidrule{3-10}\cmidrule{12-19}
Method && \multicolumn{2}{c}{Hours} && 
\multicolumn{2}{c}{Employed} && \multicolumn{2}{c}{Student} && 
\multicolumn{2}{c}{Hours} && \multicolumn{2}{c}{Employed} && 
\multicolumn{2}{c}{Student}\\
\midrule
HC$_1$, $\N(0,1)$ && 45.9 &48.8 &&61.2 &62.3 &&42.4 &43.7 
&&37.8 &26.0 &&54.3 &44.4 &&28.6 &21.7 \\
\midrule
State\tkk-year\\
CV$_{\tn1}$, $t(764)$ &&24.8 &27.5 &&30.6 &32.2 &&28.6 &29.6 
&&16.4 &16.3 &&21.1 &25.2 &&14.6 &16.2 \\
CV$_{\tn3}$, $t(764)$ &&19.3 &21.3 &&24.2 &25.9 &&23.2 &24.5
&&11.4 &13.3 &&15.4 &20.5 &&10.0 &13.7 \\
CV$_{\tn1}$, WCR &&20.6 &22.7 &&25.7 &27.1 &&25.1 &26.3 
&&13.2 &14.6 &&17.3 &22.5 &&11.8 &14.6 \\
\midrule
State\\
CV$_{\tn1}$, $t(50)$ &&7.1 &9.3 &&7.7 &10.0 &&7.0 &8.4 
&&7.5 &6.6 &&9.0 &8.3 &&6.5 &6.9 \\
CV$_{\tn3}$, $t(50)$ &&5.2 &6.6 &&5.3 &6.6 &&4.9 &6.0
&&5.5 &5.1 &&5.8 &6.0 &&4.8 &5.4 \\
CV$_{\tn1}$, WCR &&4.8 &5.7 &&4.6 &5.3 &&5.1 &5.8 
&&5.2 &4.8 &&5.3 &5.5 &&5.1 &5.4 \\
\midrule
Region\\
CV$_{\tn1}$, $t(8)$ &&7.6 &8.5 &&7.3 &8.1 &&7.8 &9.1
&&7.6 &7.2 &&8.2 &8.0 &&7.8 &7.2 \\
CV$_{\tn3}$, $t(8)$ &&4.6 &5.3 &&4.5 &5.5 &&4.9 &5.4
&&5.1 &5.0 &&5.3 &5.2 &&4.9 &4.9 \\
CV$_{\tn1}$, WCR &&5.4 &6.2 &&4.9 &5.7 &&6.0 &6.8
&&6.2 &6.0 &&6.0 &6.1 &&6.1 &5.6 \\
\bottomrule
\end{tabular*}
\vskip 6pt
{\footnotesize \textbf{Notes:} The numbers are rejection percentages
at the nominal $5\%$ level based on 10,\tkk000 simulations when a
single placebo regressor is added to \eqref{reg:wage}. For the
DiD-type treatment, the first number in each pair is the smallest
rejection percentage over all parameter values used to simulate the
placebo regressor, and the second is the largest; see text. For the
state\tkk-level AR(1) component, the first number is for $\rho=0.5,
\delta=0.9$, and the second is for $\rho=0.8, \delta=0.5$. There are
765 state\tkk-year clusters, 51 state clusters, and 9 region clusters.
The classic WCR bootstrap uses $B=999$.}
\end{table}


For the second set of simulation experiments, the placebo regressor is
generated by
\begin{equation}
x_{ist} = \delta\tkk v_{st} + (1-\delta) \epsilon_{ist},
\quad v_{st} = \rho\tkk v_{s,t-1} + e_{st}, \;\; 0 \le \rho < 1, \;\;
0 \le \delta \le 1,
\label{xprocess}
\end{equation}
where the $\epsilon_{ist}$ and the $e_{st}$ are independent standard
normals. Thus the $v_{st}$ are 51 separate stationary AR(1) processes,
and $x_{ist}$ is a weighted average of $v_{st}$ and $\epsilon_{ist}$. 
When either $\rho=0$ or $\delta=0$, the $x_{ist}$ are independent.
When both $\rho$ and $\delta$ are positive, they are correlated within
both state\tkk-years and states. They are never correlated across
states within regions. The extent to which the $x_{ist}$ are
correlated within state\tkk-years and states depends on both of the
parameters in \eqref{xprocess}. For simplicity, we report rejection
percentages for just two cases. In the first case, $\rho=0.5$ and
$\delta=0.9$, so there is a lot of correlation within each
state\tkk-year. In the second case, $\rho=0.8$ and $\delta=0.5$, so
there is less correlation within state\tkk-years but more correlation
across years within each state.




The first row in \Cref{tab:placebo} shows that failing to cluster
always leads to severe over-rejection, and the next three rows show
that clustering at the state\tkk-year level also leads to substantial
over-rejection. This is true for all methods of inference and for both
DGPs, but more so for the DiD-type one. Rows 5 through 7 show that
there can be noticeable over-rejection for clustering at the state
level when using CV$_{\tn1}$ with $t(G-1)$ critical values, especially
for the DiD-type DGP. This is to be expected given the unbalanced
cluster sizes at the state level. Both CV$_{\tn3}$ with $t(G-1)$
critical values and the classic WCR bootstrap perform much better. For
the AR(1) placebo regressors, both CV$_{\tn3}$ and the WCR bootstrap
perform very well indeed when there is clustering at either the state
or region level. For both designs, the performance of CV$_{\tn3}$ at
the region level is remarkably good. However, the WCR bootstrap tends
to outperform it slightly at the state level.




The placebo\tkk-regression results in \Cref{tab:placebo} are largely
consistent with those of the score\tkk-variance tests. They suggest
that the CV$_{\tn3}$ and classic WCR bootstrap results with
state\tkk-level clustering in \Cref{tab:teenreg} can probably be
relied upon, but that the results without clustering or with
state\tkk-year clustering should not be believed.

\section{Conclusion: A Summary Guide}
\label{sec:conc}

We conclude by presenting a brief summary guide. This is essentially a
checklist for cluster-robust inference in regression models. The first
three items should be dealt with as soon as possible in the process of
specifying and estimating a model. The remaining ones should be kept
in mind throughout the process of estimation and inference.


\begin{enumerate}

\item List all plausible clustering dimensions and levels for the data
at hand and make an informed decision regarding the clustering
structure. The decision may depend on what is to be estimated and why.
Formal tests (\Cref{subsec:level}) can be helpful in making this
decision. In some cases, placebo regressions
(\Cref{subsec:placebo,subsec:emp-placebo}) may also be informative. A
conservative approach is simply to choose the structure with the
largest standard error(s) for the coefficient(s) of interest, subject
to the number of clusters not being so small that inference risks
being unreliable.

\item For each of the plausible levels of clustering, report the number
of clusters, $G$, and a summary of the distribution of the cluster
sizes (the $N_g$). We suggest reporting at least the minimum, maximum,
mean, and median of the $N_g$. These can be reported in tabular form,
as in \Cref{tab:clustsum}. It may also be helpful to present the
distribution(s) of the $N_g$ graphically using box plots, histograms,
or EDFs like the ones in \Cref{fig:EDF}.


\item For the key regression specification(s) considered, report
information about leverage, partial leverage, and influence
(\Cref{subsec:leverage,sec:report,subsec:emplever}), including the 
effective number(s) of clusters for the coefficient(s) of interest. 
This may be particularly informative for difference\tkk-in-differences
and other treatment models. Inferences may not be reliable when a few 
clusters are highly influential or have high (partial) leverage.



\item In addition to, or instead of, the usual CV$_{\tn1}$ CRVE,
employ the CV$_{\tn3}$ CRVE and at least one variant of the restricted
wild cluster (WCR) bootstrap (\Cref{subsec:wild}) as a matter of
course for both tests and confidence intervals. In many cases,
especially when $G$ is reasonably large and the clusters are fairly
homogeneous, these methods will yield very similar inferences that can
likely be relied upon. However, in the event that they differ, it
would be wise to try other methods as well, including additional
variants of the WCR bootstrap and some of the alternative methods
discussed in \Cref{sec:other}.

\item For models with treatment at the cluster level, where either the
treated clusters or the controls are few in number and/or atypical,
cluster-robust inference can be quite unreliable
(\Cref{subsec:failure}), even when it is based on CV$_{\tn3}$ or the
WCR bootstrap. In such cases, it is important to verify that the
results are (or perhaps are not) robust. This can often be done by
using methods based on randomization inference (\Cref{subsec:RI}).

\end{enumerate}


\setlength{\bibsep}{1.5pt}
\bibliography{mnws}
\addcontentsline{toc}{section}{\refname}

\end{document}